\documentclass[entropy,preprints,accept,pdftex,moreauthors]{Definitions/mdpi} 

\firstpage{1} 
\pubvolume{1}
\issuenum{1}
\articlenumber{0}
\pubyear{2024}
\copyrightyear{2024}
\datereceived{ } 
\daterevised{ } 
\dateaccepted{ } 
\datepublished{ } 
\hreflink{https://doi.org/} 

\newcommand{\eps}{\epsilon}
\newcommand{\angstrom}{\mbox{\normalfont\AA}}
\newcommand{\photoz}{photo-$z$}

\newcommand{\snia}{SN~Ia}
\newcommand{\sneia}{SNe~Ia}
\newcommand{\snii}{SN~II}
\newcommand{\sneii}{SNe~II}
\newcommand{\gaia}{\emph{Gaia}}

\let\proglang=\textsf
\newcommand{\python}{\proglang{Python}}
\newcommand{\rlang}{\proglang{R}}

\let\swpack=\texttt
\newcommand{\npandas}{\swpack{nested-pandas}}

\DeclareMathOperator{\SP}{\mathcal{SP}}

\newcommand{\ceqn}[1]{equation~\eqref{#1}}

\newcommand{\cfig}[1]{Figure~\ref{#1}}

\usepackage[colorinlistoftodos,textsize=footnotesize]{todonotes}
\definecolor{kellygreen}{rgb}{0.3, 0.73, 0.09} 
\definecolor{palegreen}{rgb}{0.6, 0.98, 0.6}
\definecolor{hlgreen}{rgb}{0.1, .6, 0.05} 

\newif\ifenotes
\enotestrue 

\ifenotes

\newcommand{\mnote}[1]{\todo[size=\scriptsize,linecolor=kellygreen,backgroundcolor=palegreen!25,bordercolor=gray]{#1}}

\newcommand{\cnote}[1]{\todo[inline,backgroundcolor=palegreen!25,bordercolor=gray]{#1}}

\newcommand{\inote}[1]{\textcolor{hlgreen}{\textbf{\textsl{[#1]}}}}

\newcommand{\binote}[1]{\colorbox{palegreen!25}{{$\bullet\bullet\bullet$}{\sl [#1]}{$\bullet\bullet\bullet$}}}

\newcommand{\snote}[1]{\todo[linecolor=kellygreen,backgroundcolor=yellow!15,bordercolor=gray]{See \textbf{\color{dred} #1}}}

\else

\newcommand{\mnote}[1]{\relax}
\newcommand{\cnote}[1]{\relax}
\newcommand{\inote}[1]{\relax}
\newcommand{\binote}[1]{\relax}
\newcommand{\snote}[1]{\relax}

\fi

\Title{Modeling populations of light curves and spectra: \\
Bayesian functional data analysis in astronomy}

\TitleCitation{Bayesian functional data analysis in astronomy}

\Author{Thomas Loredo$^{1}$\orcidA{}*, Kaisey Mandel$^{2}$\orcidB{}, Tam\'as Budav\'ari$^{3}$\orcidC{}, Matthew Graham$^{4}$\orcidD{}, David Kent$^{5}$\orcidE{}, David Ruppert$^{6}$\orcidF{}, and Kevin Tao$^{7}$\orcidG{}}

\AuthorNames{Thomas Loredo, Tam\'as Budav\'ari, David Kent, Daniel Kowal, Kaisey Mandel, and David Ruppert}

\AuthorCitation{Loredo, T.; Budav\'ari, T.; Kent, D.; Mandel, K.; Ruppert, D.}

\address{%
$^{1}$ \quad Cornell Center for Astrophysics and Planetary Science and Department of Statistics and Data Science, Cornell University; loredo@astro.cornell.edu\\
$^{5}$ \quad Statistical Laboratory of the Department of Pure Mathematics and Mathematical Statistics, Institute of Astronomy, and Kavli Institute for Cosmology, University of Cambridge; kmandel@statslab.cam.ac.uk\\
$^{3}$ \quad Department of Applied Mathematics and Statistics, Johns Hopkins University; budavari@jhu.edu\\
$^{4}$ \quad Division of Physics, Mathematics and Astronomy, California Institute of Technology; mjg@caltech.edu\\
$^{5}$ \quad Department of Statistics and Data Science, Cornell University; dk657@cornell.edu\\
$^{6}$ \quad School of Operations Research and Information Engineering and Department of Statistics and Data Science, Cornell University; dr24@cornell.edu\\
$^{7}$ \quad Department of Statistics and Data Science, Cornell University; kt575@cornell.edu}

\corres{Correspondence: thomas.loredo@cornell.edu}

\abstract{
Cosmic demographics---the statistical study of populations of astrophysical objects---has long relied on tools from \emph{multivariate statistics} for analyzing data comprising fixed-length vectors of properties of objects, as might be compiled in a tabular astronomical catalog (say, with sky coordinates, and brightness measurements in a fixed number of spectral passbands). 
But beginning with the emergence of automated digital sky surveys, ca.~2000, astronomers began producing large collections of data with more complex structure: light curves (brightness time series) and spectra (brightness vs.\ wavelength).
These comprise what statisticians call \emph{functional data}---measurements of populations of functions.
Modern ground-based and space-based sky surveys are providing astronomers with a flood of functional data.
New methods are needed to accurately and optimally analyze large ensembles of light curves and spectra, accumulating information both along each measured function, and across a population of such functions.
\emph{Functional data analysis} (FDA) provides tools for statistical modeling of functional data.
Astronomical data presents several challenges for FDA methodology, e.g., sparse, irregular, and asynchronous sampling, heteroscedastic measurement error, and nontrivial selection effects.
Bayesian FDA uses hierarchical Bayesian models for function populations, and is well suited to addressing these challenges.
We provide an overview of astronomical functional data across many areas of astronomy, and of some key Bayesian FDA modeling approaches, including functional mixed effects models, stochastic process models, and model architectures for ``second-generation'' functional data such as dynamic spectra (bivariate spectro-temporal data).
We briefly describe two representative applications of Bayesian FDA in astronomy: a framework combining FDA and machine learning methods to build low-dimensional parametric models for galaxy spectra, and bivariate (spectro-temporal) modeling of multiband light curves from observations of Type~Ia supernovae using Gaussian processes. 
}

\keyword{astrostatistics; time series; spectroscopy; Bayesian data analysis; functional data analysis; hierarchical Bayesian modeling; Gaussian processes; dimension reduction; manifold learning}

\begin{document}



\section{Introduction}
\label{sec:Intro}

Even an hour of naked-eye observation of the night sky reveals the visible stars to be moving, relative to Earth-bound observers.
However, they move \emph{together}.
For centuries, astronomers of antiquity explained this by considering stars to comprise a population of objects fixed to a solid but invisible (or perhaps dark) celestial sphere that rotated about the Earth. 
The star-decorated sphere was known as the \emph{firmament} (from the Latin verb \emph{firmare}, ``to make firm''), populated by the so-called \emph{fixed stars}---fixed in location, and in brightness (in contrast to the few ``wanderers''---$\pi\lambda\alpha\nu\tau\acute{\epsilon}\varsigma$ [\emph{plan\"{e}t\"{e}s}] in Greek---that have unique patterns of motion, and brightness that varies in time).
Using longitude and latitude coordinates attached to the celestial sphere, each fixed star could be characterized with three numbers: the celestial longitude and latitude describing its location on the sphere, and its apparent brightness, designated by one of six magnitudes (the brightest stars were of first magnitude, the dimmest of sixth).
Astronomers compiled \emph{tabular catalogs} of these fundamental observables, as aids to navigation, and to guide ongoing study.
The oldest intact stellar catalogs date back nearly two millennia---the catalogs in Ptolemy's \emph{Almagest} and \emph{Handy Tables}, from the second century.


Beginning in the 19th century, astronomers began adapting laboratory-based measurement techniques such as spectroscopy and photography to work on telescopes.
These tools provided vastly richer measurements of diverse cosmic objects---spectra of stars and galaxies, resolved images of galaxies,
and time series (``light curves'') of brightness for variable and transient objects.
Further advances in the 20th century, across the electromagnetic spectrum, dramatically improved the precision and diversity of astronomical measurements, and provided even more complex types of data: time-varying spectra, time-varying images, and spectrally-resolved images.
In the 21st century, large-scale automated digital sky surveys began producing voluminous quantities of such \emph{sampled-function data} for cosmic populations, most of it freely accessible in public databases.
The fundamental data of most modern survey-based astronomy now comprise measurements of one-dimensional (1D), 2D, and 3D functions, rather than of a finite, fixed set of scalar observables.

It is common practice to summarize survey data with derived quantities collected in finite-width tables, qualitatively similar to those of the \emph{Almagest}.
Such tables compress the observational data (e.g., images or spectra), making catalogs of properties of large populations more manageable to share and analyze.
Statistically, tabular data enable analysis of cosmic populations using methods from \emph{multivariate statistics}---a rich and mature area offering astronomers many useful tools for analyzing tables collecting \emph{vector-valued} measurements.
But the underlying sampled-function data---images, spectra, or time series, that typically vary in size, sample spacing, and error variance across an observed population---contain significant additional information with potential to further inform astronomers' analyses of cosmic populations.

A similar change in the character of data ensembles occurred in other scientific fields in the late twentieth century, leading to the creation of a relatively new area of statistics, \emph{functional data analysis} (FDA): statistical analysis of measurements of ensembles of related functions.
Astronomers have only recently begun adapting techniques from the FDA literature to astronomy.
Emerging instrumentation will imminently produce a veritable flood of functional data, enjoining the astronomical community to more broadly use FDA across many areas of astronomy.

In this review, we begin with a broad survey of the emergence of functional data in diverse areas of astronomy, both to expose astronomers to the functional perspective on familiar datasets and data analysis problems, and to expose information scientists to the panoply of FDA opportunities across astronomy (\S~\ref{sec:Tab2Func}).
Subsequently, we provide a brief overview of FDA (\S~\ref{sec:FDA}), highlight recent FDA applications in astronomy (\S~\ref{sec:AstroFDA}), and describe two illustrative examples from recent research adopting \emph{Bayesian} approaches to FDA:
developing a low-dimensional parametric model for the population of galaxy spectra observed by the Sloan Digital Sky Survey (\S~\ref{sec:SEDPop}),
and using multiband light curves (time series) to estimate the luminosities and distances of Type~Ia supernovae, which are used to measure the cosmological expansion rate to ascertain the properties of dark matter and dark energy (SNe~Ia; \S~\ref{sec:SNIa}).
Although our focus is on FDA in astronomy, we intend this survey to be accessible to physical scientists and information scientists more generally, providing accessible examples of how FDA approaches can be relevant to the types of data that arise across many areas in the physical sciences.
A more terse overview of several of these topics is available in \cite{L+25-BFDAInAstro-short}, in the \emph{MaxEnt 2024} proceedings issue of \emph{Physical Sciences Forum}.%
\footnote{This review is based on invited presentations given by the team at the 64th International Statistics Institute World Statistics Congress in 2023, and at the 43rd International Workshop on Bayesian Inference and Maximum Entropy Methods in Science and Engineering ins 2024.}

\section{From tabular to functional data in astronomy}
\label{sec:Tab2Func}

\subsection{Tabular data over two millennia}

\begin{figure}[t]  
\begin{adjustwidth}{-.5\extralength}{0cm}
\centerline{\includegraphics[width=12 cm]{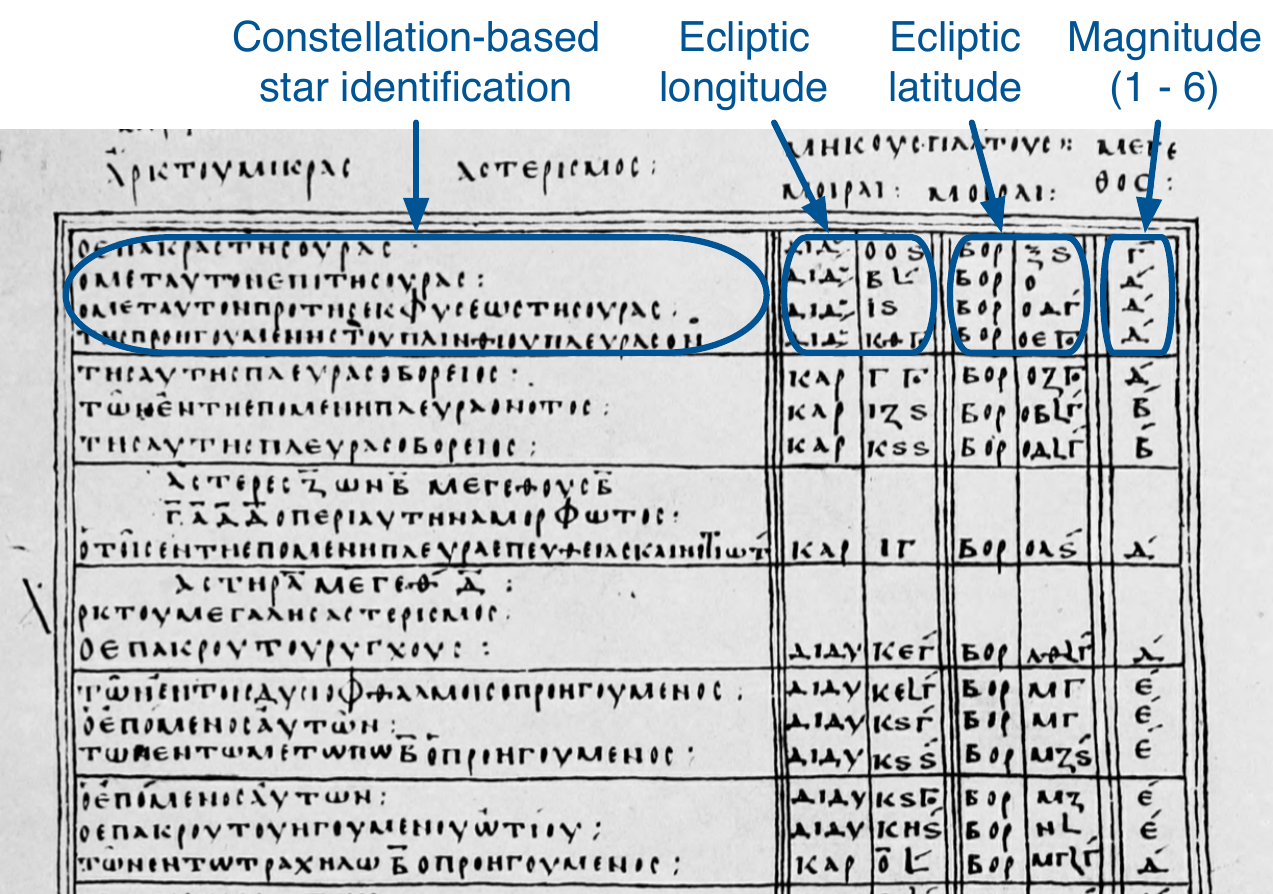}}
\caption{
The top part of a page from Ptolemy's \emph{Almagest}, annotated to identify the columns in Ptolemy's tables (image from a turn-of-the-20th-century study of \emph{Almagest} manuscripts by Peters and Knobel, \cite{KP1915-Ptolemy-Almagest}).
\label{fig:Almagest}}
\end{adjustwidth}
\end{figure}

For centuries, astronomers contemplating the properties of stellar populations relied on data collected in \emph{tabular stellar catalogues}.
As noted above, Ptolemy's second century \emph{Almagest} contains some of the earliest such tables.%
\footnote{Highlighting its immense influence, \emph{Almagest} is the Latin translation of the book's Arabic name, meaning ``The Greatest;'' Arabic translations of the Greek original circulated from the 12th to 15th centuries, when original Greek copies resurfaced.}
\cfig{fig:Almagest} shows an excerpt from a table in the \emph{Almagest}.
Each constellation had its own table; the columns contain instructions for finding a star within the constellation, the star's ecliptic longtitude and latitude, and the star's brightness as an integer magnitude, from 1 to 6, bright to dim.%
\footnote{\label{fn:mag}Ptolemy's perceptual magnitude system evolved to become the modern and somewhat peculiar astronomical apparent magnitude system, with apparent magnitude $m$ related to observed energy flux $F$ via $m = - 2.5\log(F/F_{\rm fid}$), a negative logarithmic measure relative to a fiducial flux, $F_{\rm fid}$.
There is an analogous \emph{absolute} magnitude system for luminosity, $L$ (total rate of emission of energy), with absolute magnitude $M = - 2.5\log(L/L_{\rm fid})$ for fiducial luminosity $L_{\rm fid}$.\label{fn-mag}}

\unskip



By the early twentieth century, astronomers produced many more extensive and diverse tabular catalogs, describing basic properties of stars, galaxies, nebulae, minor planets, and other families of objects.
This motivated the development of what was initially called ``statistical astronomy''---the study of \emph{populations} of astronomical objects, as groups, via the techniques of multivariate statistics and spatial statistics, i.e., techniques for modeling \emph{vector-valued data}.
Today, ``statistical astronomy'' is used more broadly, referring to any use of formal statistical methods in astronomy.
We here use \emph{cosmic demographics} specifically for statistical study of astronomical populations, typically based on survey data summarized in catalogs.
Classic examples of cosmic demography tasks and tools include:
\begin{itemize}
\item \emph{Density estimation}, e.g., estimating the distribution of apparent brightness or intrinsic luminosity of a stellar or galaxy population (the former is known as the \emph{number--size distribution} or $\log N$ -- $\log S$ distribution; the latter is called the \emph{luminosity function}).
\item \emph{Multivariate regression} (including estimating correlations, and curve fitting), e.g., for calibrating stellar and galaxy luminosity indicators (functions of observables that predict an object's luminosity).
Examples include the Leavitt law relating the period of a Cepheid variable star light curve to the star's mean luminosity, or the Tully-Fisher and fundamental plane relations, relating a galaxy's luminosity or surface brightness to observables such as rotation velocity (for spiral galaxies) or velocity dispersion (for elliptical galaxies).
\item \emph{Classification}, e.g., to distinguish stars from galaxies, to identify types of stars from features in their spectra, or types of galaxies from their morphology and color, or to distinguish types of variable and transient sources from time series data.
\item \emph{Spatial statistics}, e.g., the use of spatial correlation functions to quantify inhomogeneity in the spatial distribution of galaxies, or the use of spatial process models such as Gaussian processes or the Neyman-Scott process \cite{NS58-NSProcess} to model galaxy clustering.  
\item \emph{Structure discovery} in multivariate distributions such as clusters or sequences (reduced-dimension manifolds), via tools such as principal component analysis (PCA) or $k$-nearest neighbors.
\end{itemize}

\subsection{Functional data from automated digital sky surveys}


\begin{figure}[t]  
\begin{adjustwidth}{-.5\extralength}{0cm}
\centerline{\includegraphics[width=11 cm]{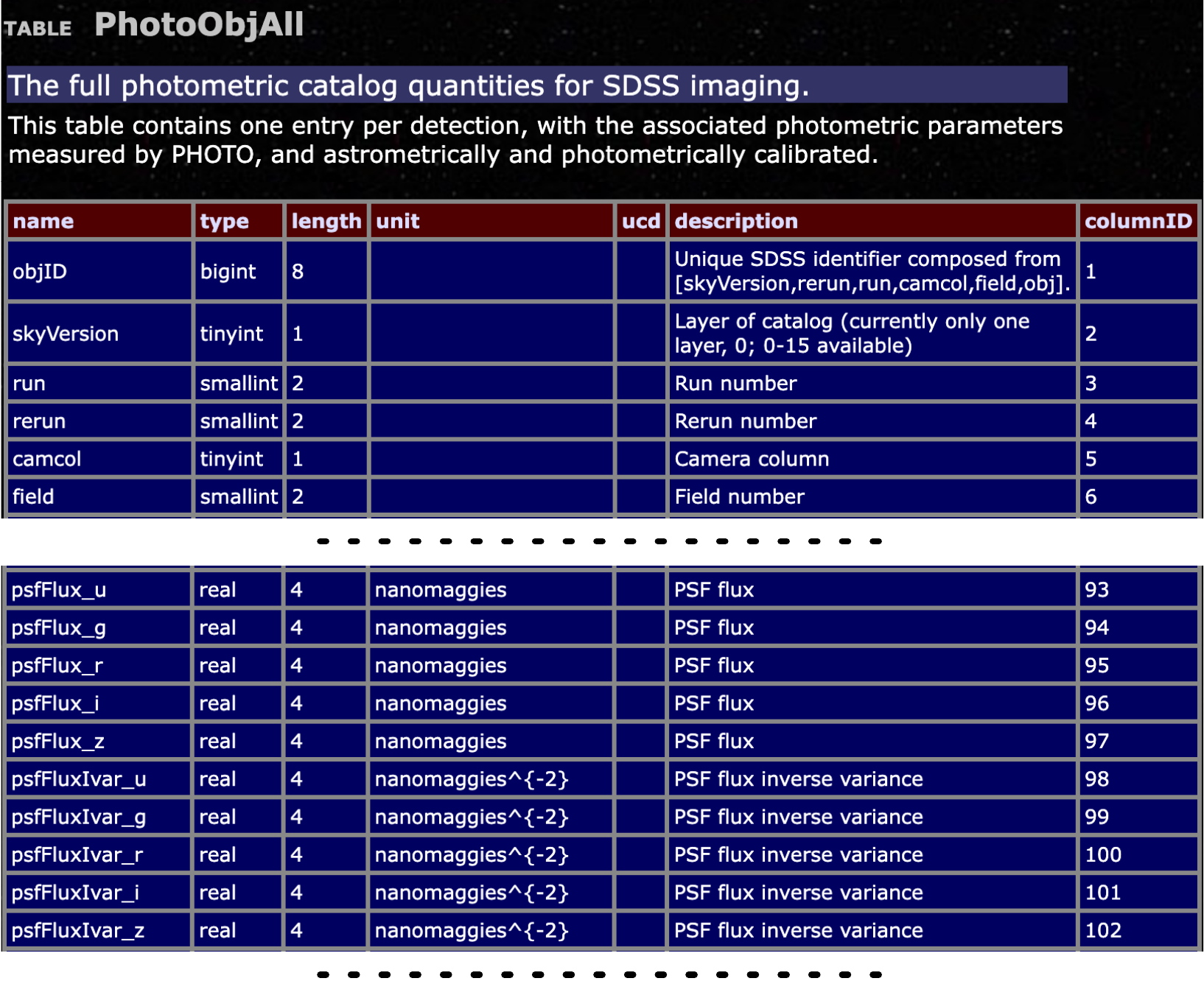}}
\caption{
Excerpts from the SQL schema describing attributes (columns) in the the \texttt{PhotoObjAll} table, one of the main tabular catalog data products of SDSS (from the SDSS SkyServer website at \url{https://skyserver.sdss.org/}, Data Release DR16).
\label{fig:SDSSSchema}}
\end{adjustwidth}
\end{figure}   
\unskip

At the turn of the twenty-first century, the first large-scale automated digital sky survey, the \emph{Sloan Digital Sky Survey} (SDSS), began producing vast tabular catalogs, dwarfing the scale of previous catalogs.
For example, the SDSS \texttt{PhotoObjAll} table reports measurements of nearly a half-billion stars and galaxies imaged in the northern sky, in a table with over 500 columns, accessible to astronomers via an SQL-like database.
\cfig{fig:SDSSSchema} shows just part of the SQL schema for the \texttt{PhotoObjAll} table.
Among the columns are coordinate and magnitude measurements (\emph{photometry} in astronomical parlance) akin to those of the \emph{Almagest}, albeit far more precisely defined and measured (and, importantly, supplemented with \emph{quantifications of uncertainty}, provided via measurement error inverse variances).
Magnitude measurements are available in various \emph{spectroscopic bandpasses} (``filters'' or ``passbands,'' indicated by the \emph{u, g, r, i}, and \emph{z} subscripts in the schema,  denoting bands that roughly cover ultraviolet, green, red, and two infrared wavelength intervals), providing quantification of the colors of objects.
The measurements also include low-dimensional parametric descriptions of the shapes of galaxies, derived from the underlying spatially-resolved image data.


Importantly, besides its tabular photometric catalogs, SDSS produced other large collections of non-tabular data: spectroscopic and image databases.
These databases report measurements of \emph{functions}: one-dimensional spectra (observed light flux density vs.\ wavelength) in spectroscopic databases, and two-dimensional images (flux density vs.\ camera focal plane coordinates) in image databases.
Of course, such measurements long predate SDSS; e.g., quantitative spectroscopy transformed astronomy to astrophysics beginning in the late nineteenth century \cite{HK84-HistAstroH2H,H10-SpecHist}.
But SDSS made large samples of calibrated spectra and images readily available to the astronomical community for the first time, enabling demographic modeling of spectrum and image ensembles at an unprecedented scale.

Where classical cosmic demographics viewed the sky as full of \emph{populations of vectors}, large-scale digital sky surveys like SDSS provide cosmic demographers access to a sky full of \emph{populations of functions}.
In the remainder of this section, we present examples of functional data in astronomy, to visually and qualitatively convey the nature of such data and the inference problems they entail.%
\footnote{The emerging area of \emph{exploratory functional data analysis} develops tools for exploration of functional data via visualization, data summarization or compression, and other techniques; see \cite{Q+24-EFDA}.}
Subsequent sections describe key FDA approaches that can address such problems.

\subsection{Stellar spectrum populations}

\begin{figure}[ht!]  
\begin{adjustwidth}{-.2\extralength}{0cm}
\includegraphics[width=14.5 cm]{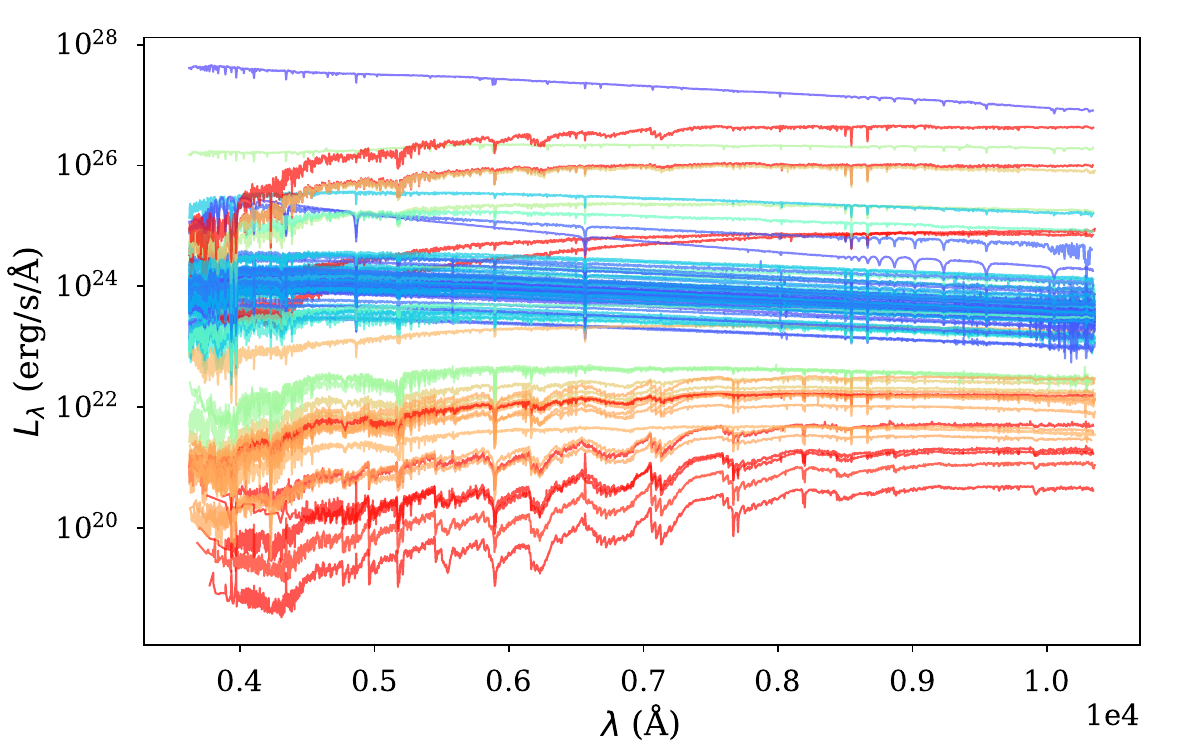}
\caption{
One hundred representative stellar spectra from the MaStar collection of stellar spectra observed by SDSS.
Line color conveys whether a particular spectrum is weighted toward short, middle, or long wavelengths, in a spectral sequence (blue to red); see text for details.
\label{fig:MaStarLum}}
\end{adjustwidth}
\end{figure}

\cfig{fig:MaStarLum} shows 100 representative spectra from a \emph{population of stellar spectra} observed by SDSS.
These 100 spectra are a subset of over 24,000 spectra comprising the MaStar catalog, whose selection criteria target capturing a diverse sample of stars whose distances are known via measurement of geometric parallax by the \gaia\ astrometric space observatory.%
\footnote{``MaStar'' reflects the origin of this stellar spectrum library from images obtained by the SDSS Mapping Nearby Galaxies at Apache Point Observatory (MaNGA) survey.
Data used here were obtained from \url{https://www.sdss4.org/surveys/mastar/}.}
Knowing the distance enables use of the inverse square law to convert the apparent flux density (energy per unit time, area, and wavelength reaching Earth) to the luminosity density (energy per unit time and wavelength emitted by the star, assuming isotropic emission; corrections for absorption and scattering of light along the line of sight were also made).
The curves displaying the spectra are colored according to the location of the peak in a smoothed version of the spectrum, with colors in spectral sequence (red corresponding to peaks at the long wavelength end of the spectrum, and blue corresponding to peaks at the short wavelength end).


Simple visual examination of the plot offers several interesting insights:
\begin{itemize}
\item The spectra are not ``completely random;'' there are visually evident patterns in the population of spectra that one would hope to be able to describe quantitatively via some kind of probability distribution over spectra.
\item Stars span over six orders of magnitude in luminosity, a huge range.
\item Red stars tend to be dim, though a subset of red stars are exceptionally bright (these are red giants, old, evolved stars nearing the ends of their lives).
\item Excepting the red giants, the bluer a star is, the more luminous it tends to be.
\end{itemize}

\unskip

\begin{figure}[H]  
\begin{adjustwidth}{-.2\extralength}{0cm}
\includegraphics[width=14.5 cm]{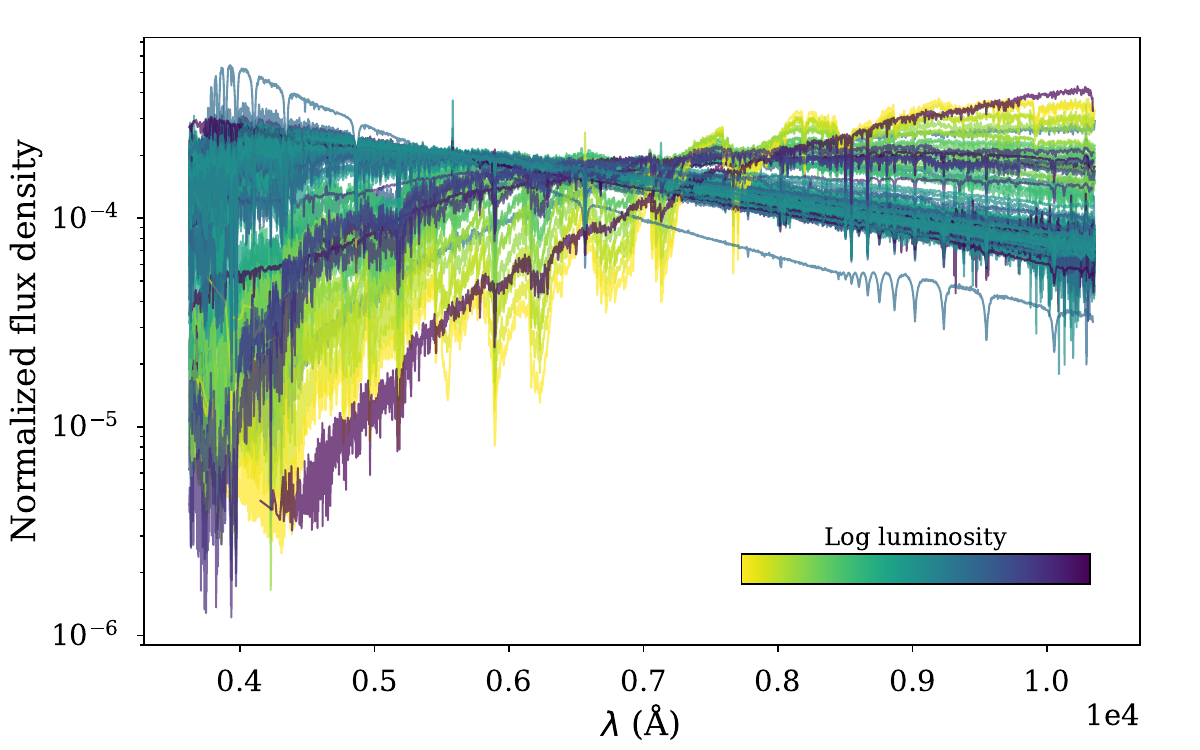}
\caption{
The 100 spectra of Figure~\ref{fig:MaStarLum}, here displayed as \emph{normalized} spectra to better display the diverse shapes of stellar spectra.
The spectra are colored according to the (logarithmic) luminosity of the star (in the observed wavelength range), as indicated by the colorbar (with luminosity increasing from left to right).
\label{fig:MaStarNorm}}
\end{adjustwidth}
\end{figure}   
\unskip


\medbreak

The huge range in luminosities spanned by the spectra makes it somewhat challenging to appreciate the diversity of shapes of the spectra in \cfig{fig:MaStarLum}.
To make the diversity more apparent, \cfig{fig:MaStarNorm} shows the same set of spectra as \emph{normalized} spectra, with the luminosity density divided by its integral (over the observed range).
Here the spectra are colored by the luminosity.
The low to moderate luminosity spectra (colored yellow to green) appear to form a smooth sequence, starting with red spectra (peaking to the right, at large wavelengths) and moving to bluer spectra.
On the other hand, the high luminosity spectra (purple colors) show diverse shapes, including some very red spectra, a cluster of spectra more moderately red, and some very blue spectra (this high-luminosity part of the distribution of spectra is evidently poorly sampled in a set of just 100 spectra).


Figures \ref{fig:MaStarLum} and \ref{fig:MaStarNorm} together suggest it could be useful to ``factor'' the distribution of spectra into a product of a distribution for overall luminosity (i.e., a stellar luminosity function), and a conditional distribution for spectrum shape given luminosity, or vice versa.
FDA addresses such factorizations via various \emph{functional regression} approaches, described below.

\begin{figure}[t!]  
\begin{adjustwidth}{-.3\extralength}{0cm}
\includegraphics[width=15 cm]{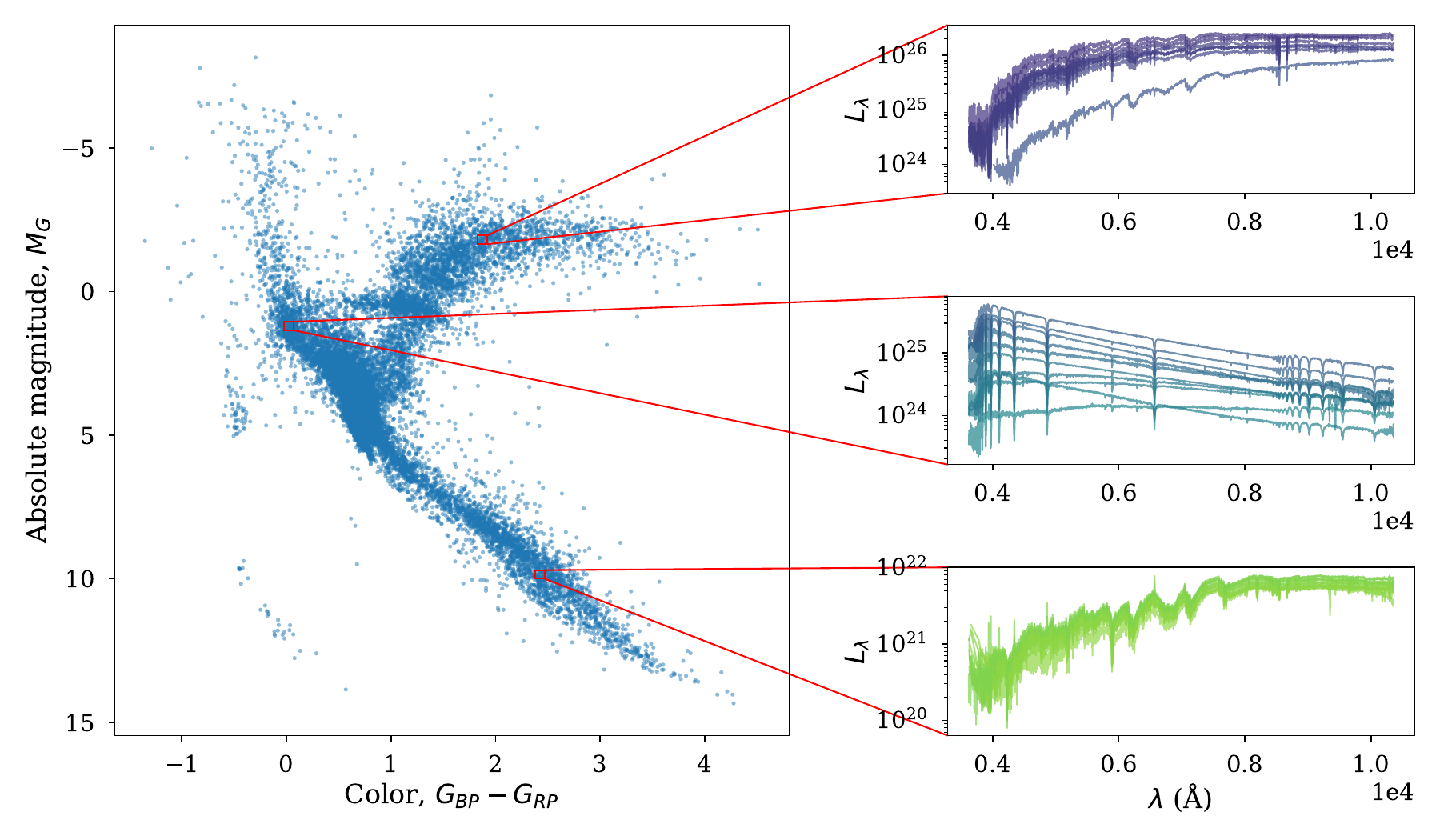}
\caption{
A color-magnitude diagram (CMD) for the MaStar sample (left panel), as a scatterplot of absolute magnitude in the \gaia\  $G$ band vs.\ $G_{BP} - G_{RP}$ color using apparent magnitudes in the \gaia\  $BP$ and $RP$ bands.
Right panels show spectra (unnormalized, i.e., luminosity density) of stars in the regions indicated by the connected small red boxes on the CMD, with spectra colored by total lumosity using the color map of \cfig{fig:MaStarNorm}.
\label{fig:MaStarCMD}}
\end{adjustwidth}
\end{figure}   
\unskip

Astrophysically, the structure in these plots reflects well-known features apparent in \emph{Hertz\-sprung-Russell (HR) diagrams}, an umbrella term for scatterplots showing stellar luminosity (intrinsic brightness) vs.\ some spectrally-derived measure of color or temperature.
When luminosity is measured by absolute magnitude (a negative-logarithmic measure of luminosity; see footnote~\ref{fn-mag}), and color by the difference of apparent magnitudes in two passbands (i.e., a negative-logarithmic measure of a flux ratio), such a diagram is called a \emph{color-magnitude diagram} (CMD).


\cfig{fig:MaStarCMD} shows a CMD (left panel) for the MaStar sample, with luminosity measured by the absolute magnitude in the \gaia\ $G$ (``green'') band, $M_G$, and color by the difference between \gaia\  blue and red band apparent magnitudes, $G_{BP} - G_{RP}$ (smaller values indicating bluer stars).
The vast majority of stars lie close to a Y-shaped locus or band in the CMD.
The right panels show collections of spectra from three small regions on this locus, revealing that spectra with similar CMD coordinates are similar in shape and luminosity.
The CMD reveals that salient features of stellar spectra can be captured by just two ``coordinates.''
This reflects the observation in the first bullet point above---stellar spectra, viewed as a function ensemble, have limited diversity.
The CMD indicates that stellar spectra lie near a low-dimensional manifold in the space of possible intensity functions.
Notably, one of the coordinates on the manifold---color---is a nonlinear functional of the luminosity density (luminosity vs.\ wavelength); it depends on a ratio of linear functionals, the magnitudes in the two bands.

The discovery of this structure in H-R diagrams around the turn of the twentieth century was hugely important, guiding the development of the astrophysical theory of stellar structure and evolution \cite{HK84-HistAstroH2H}.
Roughly speaking, the mass and age of a star largely determine where on the locus the star lies, with chemical composition being an additional important property largely responsible for the width of the locus.
The astrophysical importance of such structure in the distribution of stellar spectra highlights the potential impact of statistical methodology for discovering structure in function ensembles.

\subsection{Galaxy spectrum populations}

\begin{figure}[t]  
\begin{adjustwidth}{-.5\extralength}{0cm}
\centerline{\includegraphics[width=11 cm]{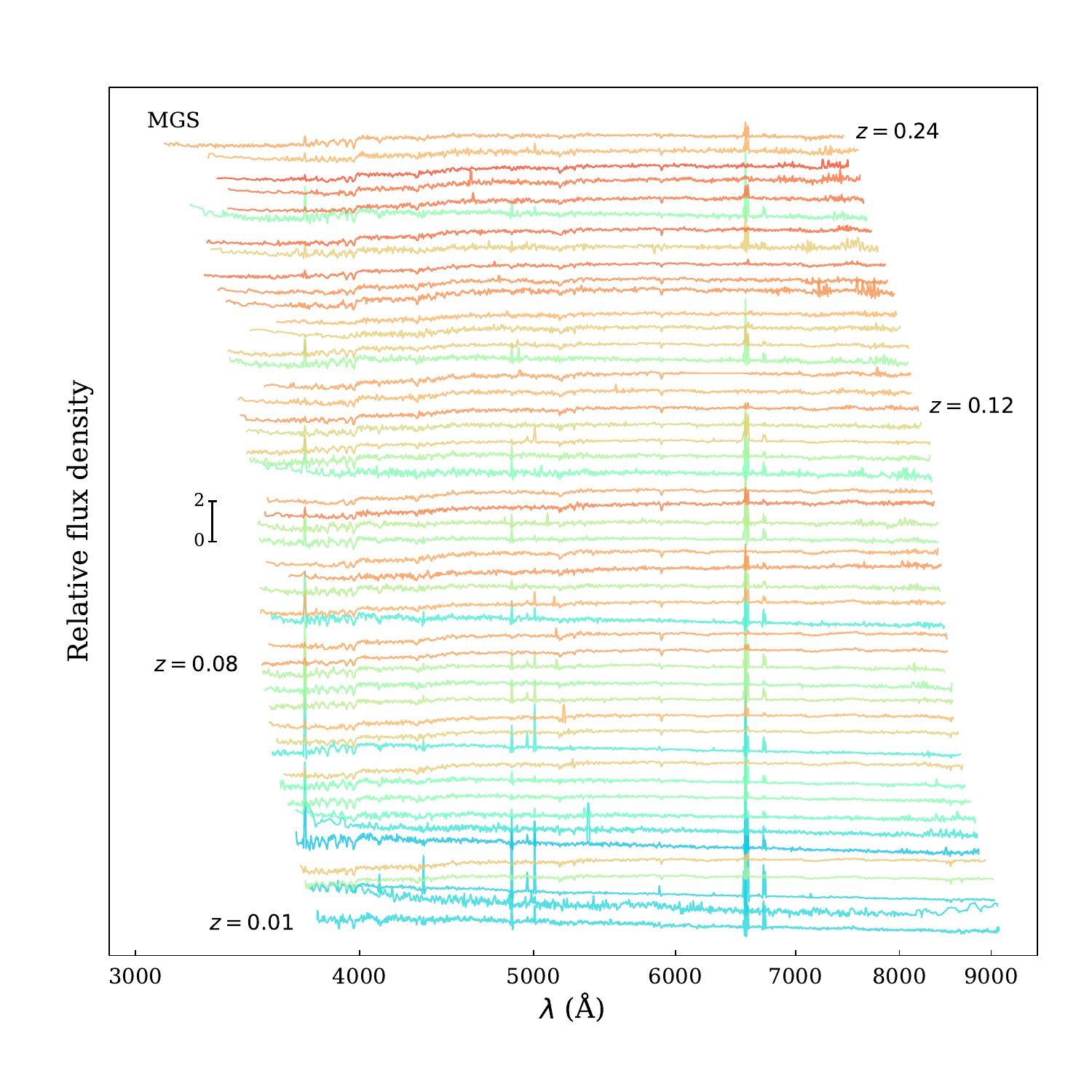}}
\caption{
One hundred representative galaxy spectra from the SDSS Main Galaxy Sample (MGS), which contains spectrophotometrically calibrated spectra for $\approx 6.8\times 10^{5}$ galaxies.
The curves show the rest-frame flux density relative to the flux density at 5500~\angstrom, but with a sequence of equal vertical coordinate offsets so the many spectra may be distinguished.
The spectra are ordered by redshift, $z$, from bottom (lowest redshift) to top; four redshift values are annotated.
The vertical bar on the left, spanning 0 to 2, shows a representative flux density ordinate span for a single spectrum.
The color of a curve is a visual indicator of the intrinsic color of the spectrum; see text for details.
\label{fig:MGS}}
\end{adjustwidth}
\end{figure}   
\unskip

Shifting focus from stars to galaxies, \cfig{fig:MGS} shows 100 representative spectra from a \emph{population of galaxy spectra} observed by SDSS.
These 100 spectra are a subset of over $\approx$~680,000 spectra comprising the SDSS \emph{Main Galaxy Sample} (MGS) \cite{S+02-SDSS-MGS}.
The cosmological expansion of the universe causes the observed spectrum of a galaxy to be redshifted vs.\ the intrinsic, rest-frame spectrum.
Via the locations of spectral lines in the spectra, the \emph{spectroscopic redshift}, $z$, is estimated for each galaxy; thanks to the Hubble law (relating a galaxy's recession velocity to distance), redshift can serve as a proxy for distance, a key galaxy property.
For each galaxy, the observed flux density, $F_{\rm obs}(\lambda_{\rm obs})$, has been converted to the (relative) flux density in the galaxy's rest frame, $F(\lambda) \propto F_{\rm obs}(\lambda\times(1+z))$.
The figure shows spectra on a relative scale, stacked in order of increasing redshift from bottom to top.
The color of a curve is a visual indicator of the intrinsic color of the spectrum (relative to the observed rest-frame wavelength range for each galaxy).
It is assigned based on the spectrum-weighted mean wavelength for each spectrum, using a rainbow-like color map (red to violet): spectra with mean wavelengths toward the large-wavelength end of their observed range are assigned redder colors, and those with mean wavelengths toward the small-wavelength end of their range are assigned bluer colors.
The color map was chosen to be applicable across this figure and two subsequent figures (the MGS sample shown here does not include spectra at the violet end of the map).

A consequence of cosmological redshift with important consequences for data analysis is that the pixel-based measurements comprising observed spectra of galaxies are \emph{not aligned in wavelength}, and the spectra necessarily do not cover the same rest frame wavelength ranges (see \S~\ref{sec:SEDPop} for further illustration of this point).

The SDSS telescope is used for both imaging surveys (producing photometric catalogs, with star and galaxy positions and magnitudes) and for spectroscopic surveys (which can target only a small fraction of the detectable stars and galaxies).
The MGS sample was selected using apparent magnitude and surface brightness thresholds, chosen to limit the number of galaxies in the sample to the number that SDSS could obtain spectra for during the early years of SDSS operations.
The thresholds impact how likely it is for different types of galaxies to be in the sample, as a function of redshift (distance).
Separate criteria were devised to obtain \emph{two companion samples} targeting particularly interesting galaxy populations that are rare in MGS.

\begin{figure}[t]  
\begin{adjustwidth}{-.5\extralength}{0cm}
\includegraphics[width=8.1 cm]{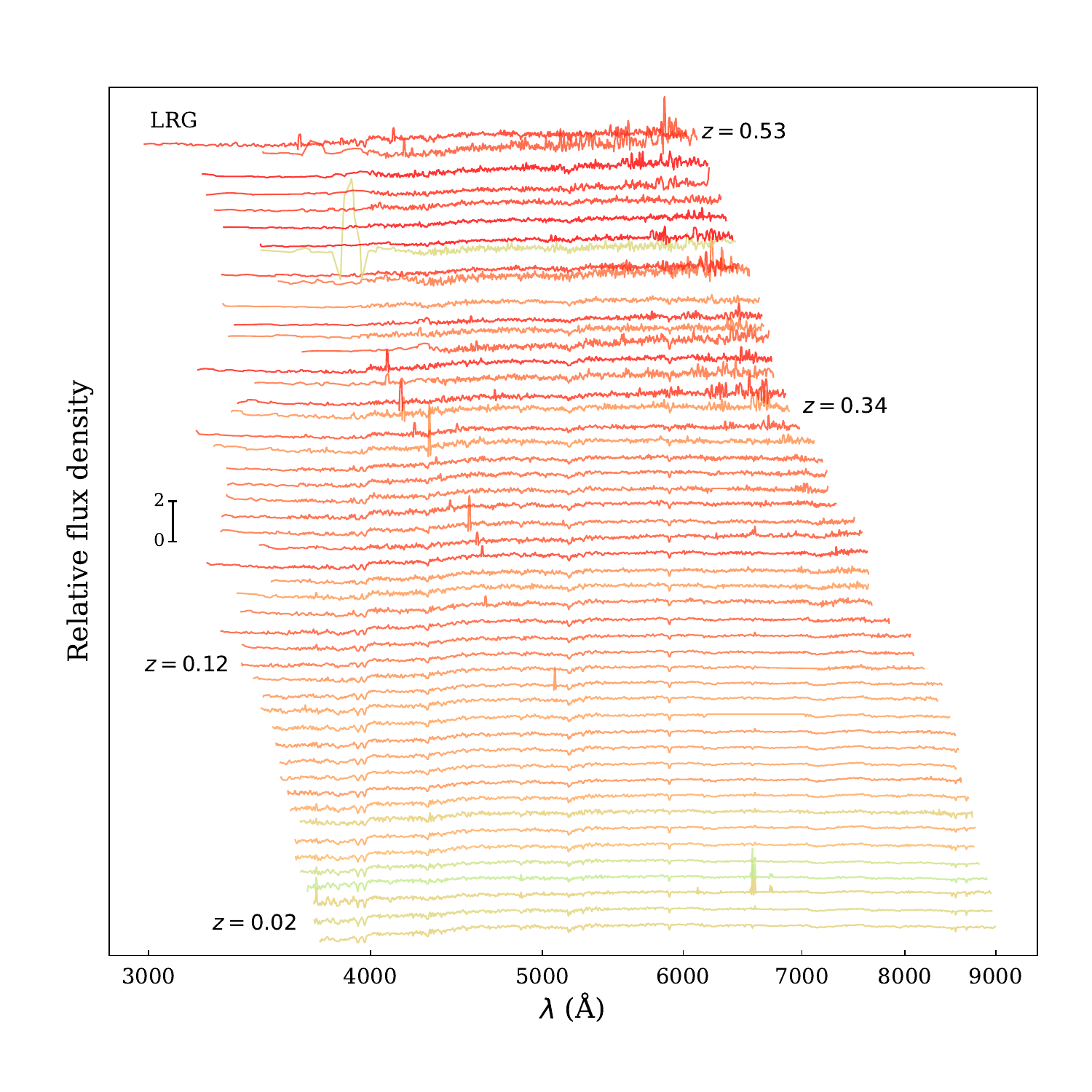}
\includegraphics[width=8.1 cm]{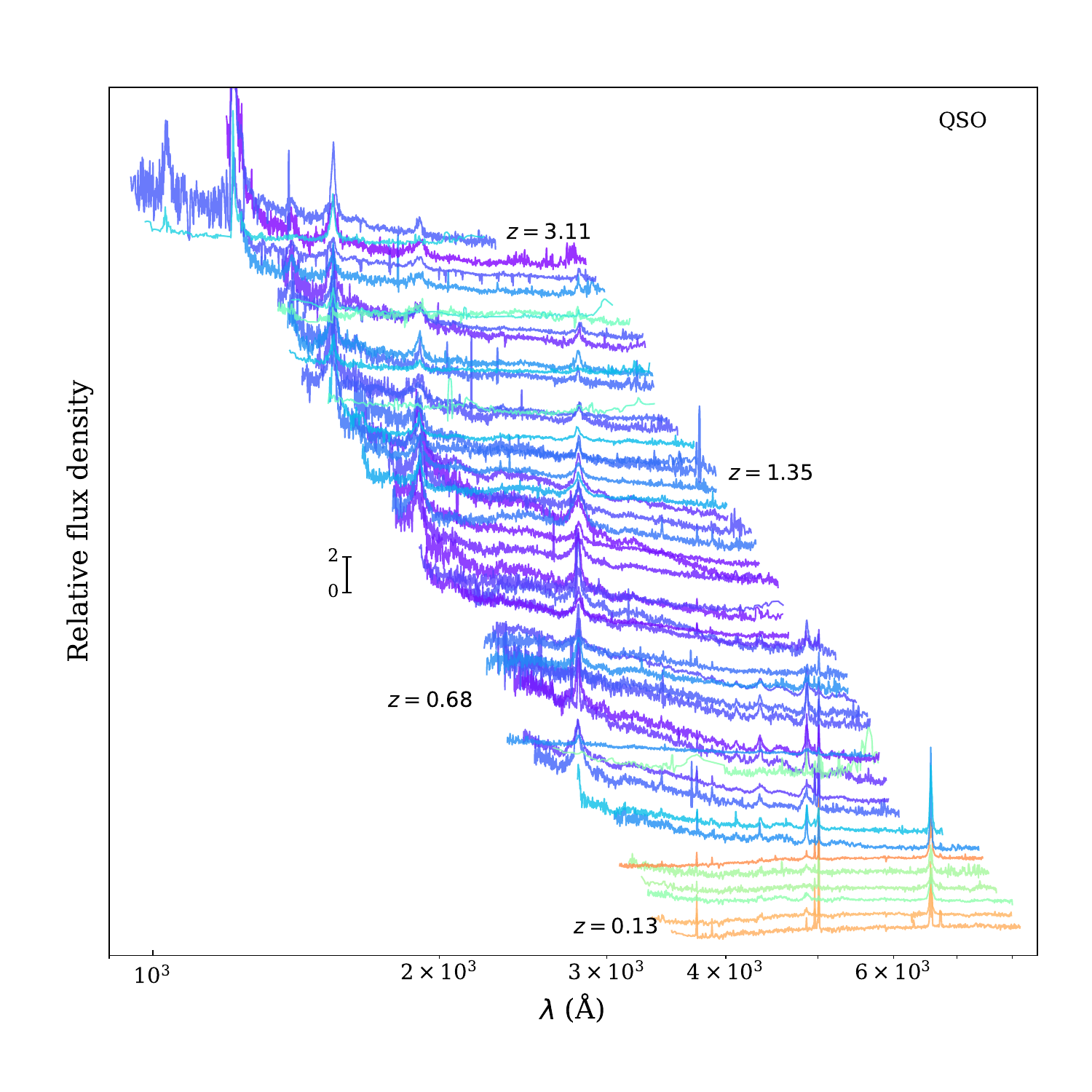}
\caption{
As in \cfig{fig:MGS}, but for different galaxy samples:
\emph{Left:} 100 representative galaxy spectra from the SDSS LRG catalog.
\emph{Right:} One hundred representative galaxy spectra from the SDSS QSO catalog.
\label{fig:LRG+QSO}}
\end{adjustwidth}
\end{figure}

The left panel of \cfig{fig:LRG+QSO} shows 100 representative galaxy spectra from the SDSS Luminous Red Galaxy (LRG) catalog, which contains spectrophotometrically calibrated spectra for  $\approx 1.9\times 10^{5}$ LRGs \cite{E+01-SDSS-LRG}.
LRGs are unusually large galaxies with old stellar populations; they comprise an exceptionally bright and relatively homogeneous population, making them useful for cosmology (since they can be seen to great distances).
Comparing with \cfig{fig:MGS}, the red character of this population is evident, and the high luminosity of LRGs allows this sample to probe to redshifts twice that of the MGS.

The right panel of \cfig{fig:LRG+QSO} shows 100 representative galaxy spectra from the SDSS Quasi-Stellar Object (QSO) catalog, which contains spectrophotometrically calibrated spectra for $\approx 7.4\times 10^{4}$ QSOs \cite{R+02-SDSS-QSO}.
A QSO is a galaxy whose emission is nearly completely dominated by light from an active galactic nucleus (AGN)---a supermassive black hole at the galaxy's center that is actively accreting gas and stars from its immediate vicinity---so its image is star-like.
QSOs are also known as \emph{quasars} (for ``quasi-stellar radio source''), particularly when discovered via radio observations.
This population has very blue spectra (i.e., dominated by short-wavelength light); the light is dominated by accretion activity associated with the supermassive central black hole, rather than coming from normal stellar populations.

\unskip



As was the case for stellar spectra, above, it is visually evident that galaxy spectra are not arbitrarily diverse.
Also, the shapes of spectra of LRGs and QSOs are systematically different from those of the typical galaxies predominant in the MGS catalog.
This raises the question of whether there is a galaxy analog to the H-R diagram---i.e., a small set of parameters that captures the salient diversity among galaxy spectra---and whether subpopulations may occupy different regions in such a parameter space.
Astrophysical modeling of galaxy spectra considers them to arise from a superposition of spectra of stellar populations, with a possible AGN component.
Such models typically have a dozen or more physical parameters, describing the distribution of stellar masses, ages, and chemical compositions, their spatial distribution, and physical characteristics of the AGN.
Fitting such stellar population synthesis (SPS) models to observed spectra reveals significant degeneracies (nonidentifiability) among the parameters, empirically motivating the hypothesis that the physically realized distribution of spectra has a lower effective dimension than the number of parameters needed to describe a galaxy a priori.
Revealing low-dimensional manifold structure in the distribution of galaxy SEDs could help guide theoretical modeling of galaxy spectra.
It could also guide construction of data-driven empirical models for galaxy spectra that could be useful for simulating survey observations, and other astrophysical tasks.

We can view galaxy spectra as a function population, and use tools from FDA to uncover structure in the distribution of spectra.
We describe an FDA approach to modeling the population of galaxy spectra in \S~\ref{sec:SEDPop}.

\subsection{Light curve (time series) populations}

The so-called fixed stars are in fact not fixed in brightness.
Every star, when observed repeatedly with sufficient photometric precision, has a time-variable flux (e.g., due to stellar activity---star spots, prominences, flares, etc.).
For most stars, the variability is well below the one-percent level.
To the naked eye (and even with good-quality modern instrumentation), virtually all stars appear fixed in magnitude; significant variability is a rarity among the visible stars.
But as instrumentation provides more precise measurements of visible stars, and, especially, measurements of increasingly vast numbers of stars and galaxies not visible to the naked eye, relatively rare objects with strong variability are seen in large numbers.
These include periodic variable stars, flaring stars, and explosive transients such as novae and supernovae seen in optical observations, or gamma ray bursts (GRBs) and fast radio bursts (FRBs) in the gamma ray and radio bands.
Even galaxies can vary in brightness, when a significant part of their emission is produced by an AGN whose accretion rate is varying with time.

One of the earliest naked-eye variable stars to be discovered is Mira, a star in the constellation Cetus that fades from roughly 2nd magnitude to invisibility and back with a period of $\approx$~332 days.
It is not a coincidence that its name is from the same Latin root from which ``miracle'' derives; its strong and periodic variability shocked the 17th-century astronomers who first realized it varied strongly and regularly.
It is the prototype for the class of long-period variable stars known as Mira variables.

The naked-eye variable $\delta$~Cephei is also a periodic variable star, varying by about one magnitude, with a period of $\approx$~5.4 days.
It is the prototype for the \emph{Cepheid variable} class of periodic variable stars that play a crucial role in constructing the cosmic distance scale.

\begin{figure}[t]  
\begin{adjustwidth}{-.5\extralength}{0cm}
\includegraphics[width=7.85 cm]{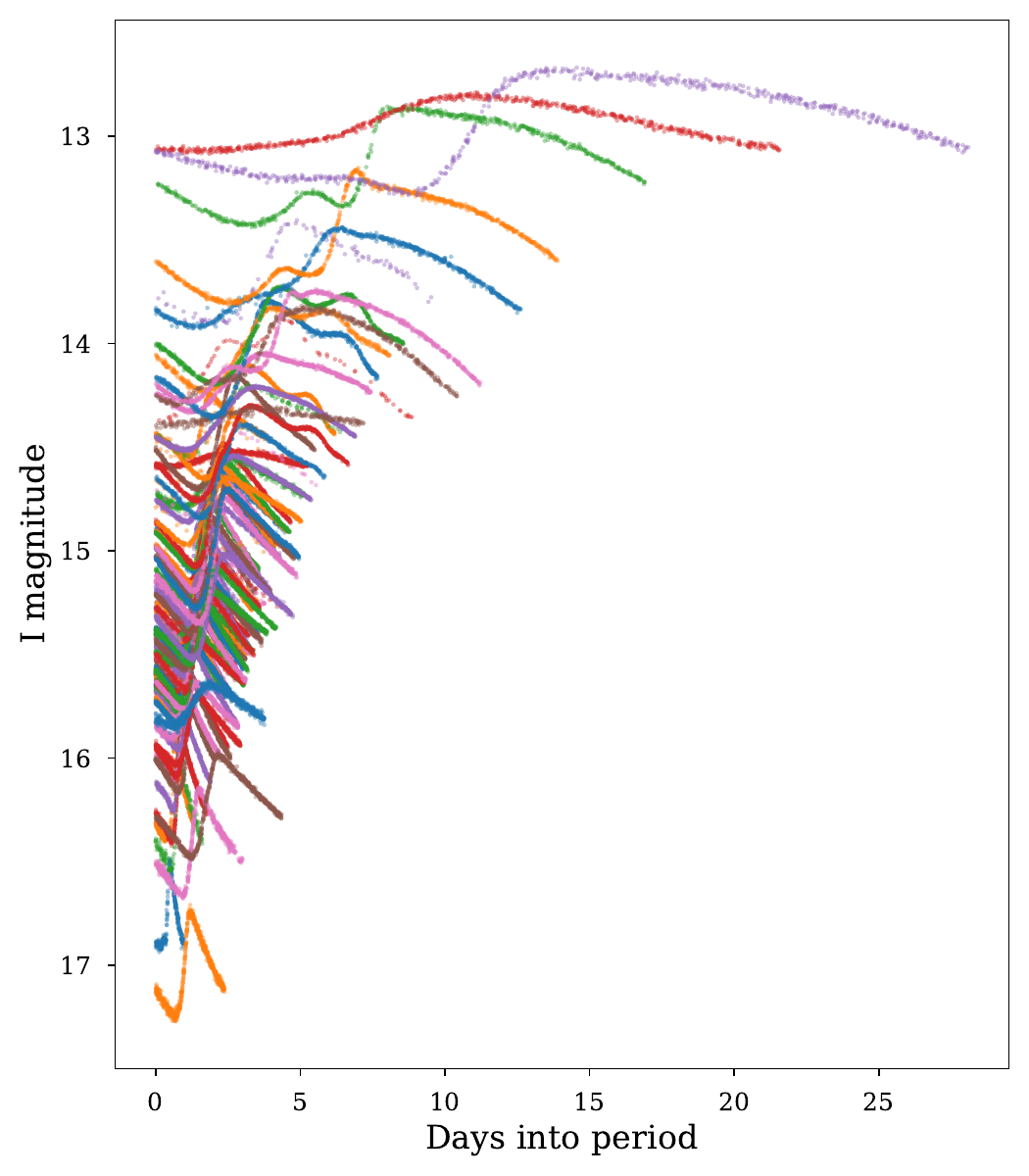} \quad
\includegraphics[width=7.85 cm]{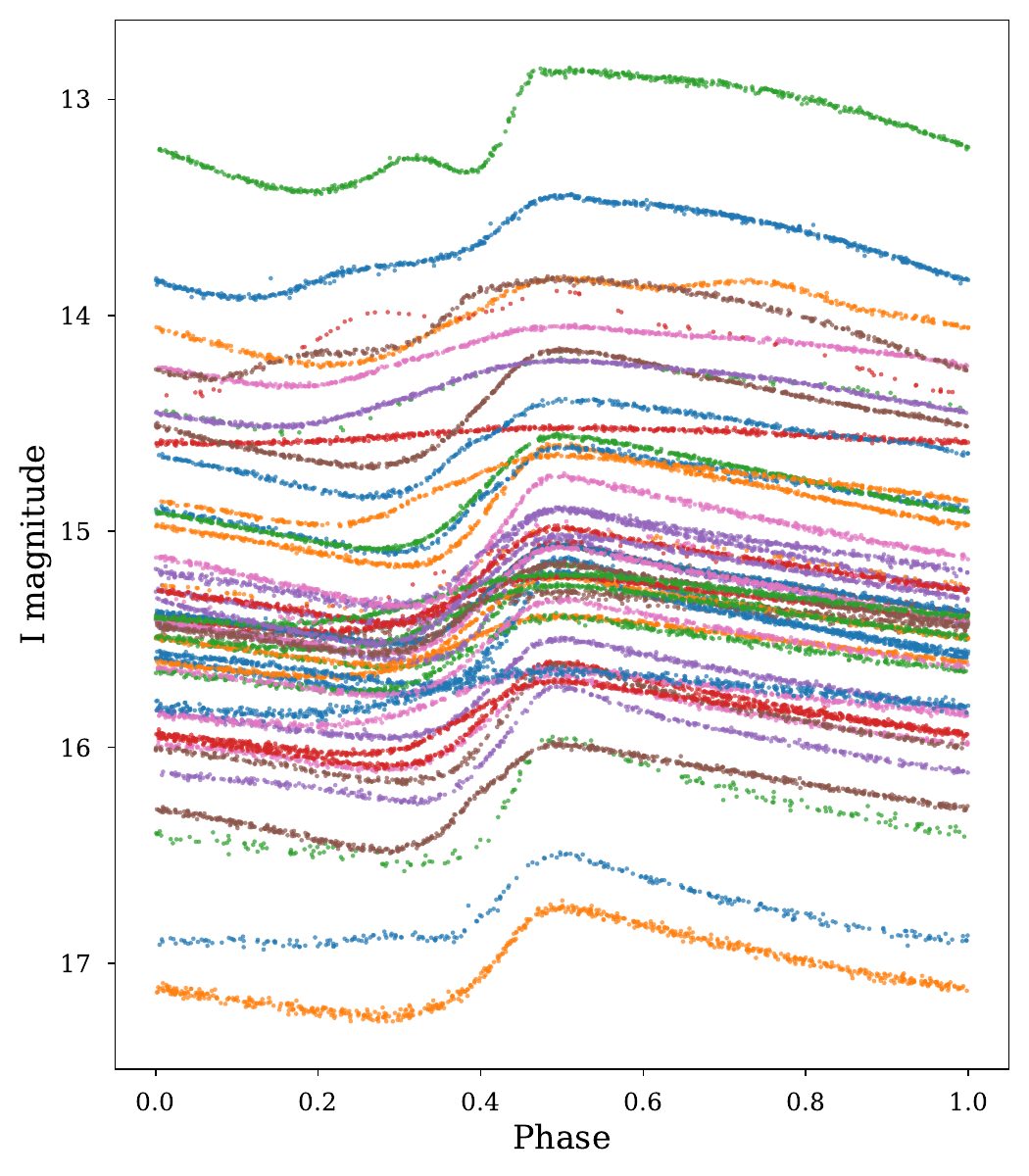}
\caption{
\emph{Left:} 103 representative light curves (apparent magnitude vs.\ time) for fundamental-mode classical Cepheid variable stars in the Large Magellanic Cloud (LMC), using OGLE~IV data from the \href{https://ogle.astrouw.edu.pl/main/collections.html}{OGLE Collection of Variable Stars (OCVS) data archive}.
(The light curves are colored in an arbitrary color cycle to aid distinguishing them.)
Observations are folded in time modulo each Cepheid's period (determined via Fourier methods).
The OCVS contains observations of over 4700 classical Cepheids in the LMC, nearly 2500 of them fundamental-mode Cepheids.
\emph{Right:} Half of the light curves from the left panel, now plotted as a function of phase with respect to the measured period for each star's pulsations, to display the diversity of Cepheid light curve shapes, and possible shape feature trends.
\label{fig:LMCCepheids}}
\end{adjustwidth}
\end{figure}

\cfig{fig:LMCCepheids} shows 103 representative light curves (apparent magnitude time series) for Cepheid variables observed in the Large Magellanic Cloud (LMC) by OGLE~IV, the fourth survey undertaken by the Optical Gravitational Lensing Experiment (OGLE) \cite{S+17-OGLEIV-LMCCepheids}.
The LMC is a small, irregular companion galaxy to the Milky Way, at a distance of $\approx$~50~kpc (160,000 light years).
Because all of the Cepheids in this sample are in the LMC (and thus at nearly the same distance), the apparent magnitude provides a measure of relative Cepheid luminosity, and the left panel reveals that long-period Cepheids are more luminous than short-period Cepheids.
It is this property that makes them so important---measuring the period allows estimation of the luminosity; combining this with measurement of the apparent magnitude, and using the inverse-square law, enables estimation of the distance to a Cepheid.
Cepheids are giant stars, readily visible in other nearby galaxies.
The Cepheid period-luminosity relationship---known as the Leavitt law, after its discoverer, Henrietta Leavitt \cite{BHR25-LeavittLegacy}---thus enables measurement of the distances to nearby galaxies (here ``luminosity'' refers to time-averaged absolute magnitude).

\unskip

Fundamentally, the task of mapping from observations of a Cepheid light curve to the Cepheid luminosity is an example of \emph{scalar-on-function regression}---using measurements of a function (the light curve) to predict a scalar (luminosity).
The conventional way astronomers accomplish this is to measure the period and mean (time-averaged) magnitude from the light curves of Cepheids at known distances to calibrate the period-luminosity law.
Of course, there is potentially more information in the light curve measurements than is captured in those two scalar summaries.
The right panel of \cfig{fig:LMCCepheids} shows the same light curve measurements displayed in the left panel (for half of the light curves, for clarity), but now plotted as a function of phase (in the $[0,1]$ interval) over the Cepheid period, rather than vs.\ time.
These curves focus attention on the light curve \emph{shapes}.
Astronomers would like to discover and exploit structure in the distribution of Cepheid light curve shapes to potentially improve prediction of Cepheid luminosities.

Periodic variable stars are persistantly variable.
There are also \emph{transient} astrophysical phenomena, producing one or more prominent bursts of emission at unpredictable times.
The most spectacular of these in the visible spectrum are supernovae, massive explosions occuring late in the evolution of stars.
Historically, supernovae were initially classified spectroscopically, via the presence or absence (and evolution) of various spectral lines.
Physically, there are two main types of supernovae: core-collapse supernovae, produced by the gravitational collapse of a massive star that has exhausted the nuclear fuel in its core, and thermonuclear supernovae, caused when interaction of a white dwarf star with a companion star causes runaway nuclear fusion, explosively disrupting the white dwarf.

We focus here on two spectroscopic classes: \emph{Type~Ia supernovae} (\sneia), comprising thermonuclear supernovae, and \emph{Type~II supernovae} (\sneii), comprising several subclasses of core-collapse supernovae (reflecting different properties of the massive progenitor stars).
Spectroscopic observations are time-consuming and only feasible for relatively bright sources.
As a result, many supernovae have only photometric measurements---irregularly sampled light curves in a few spectral bandpasses.
Modern time-domain photometric survey campaigns are detecting far more supernovae than could possibly be measured spectroscopically.
Astronomers would like to extract as much information from multiband supernova light curves as possible, for classification of supernovae and other tasks.

\begin{figure}[t]  
\begin{adjustwidth}{-.5\extralength}{0cm}
\includegraphics[width=15 cm]{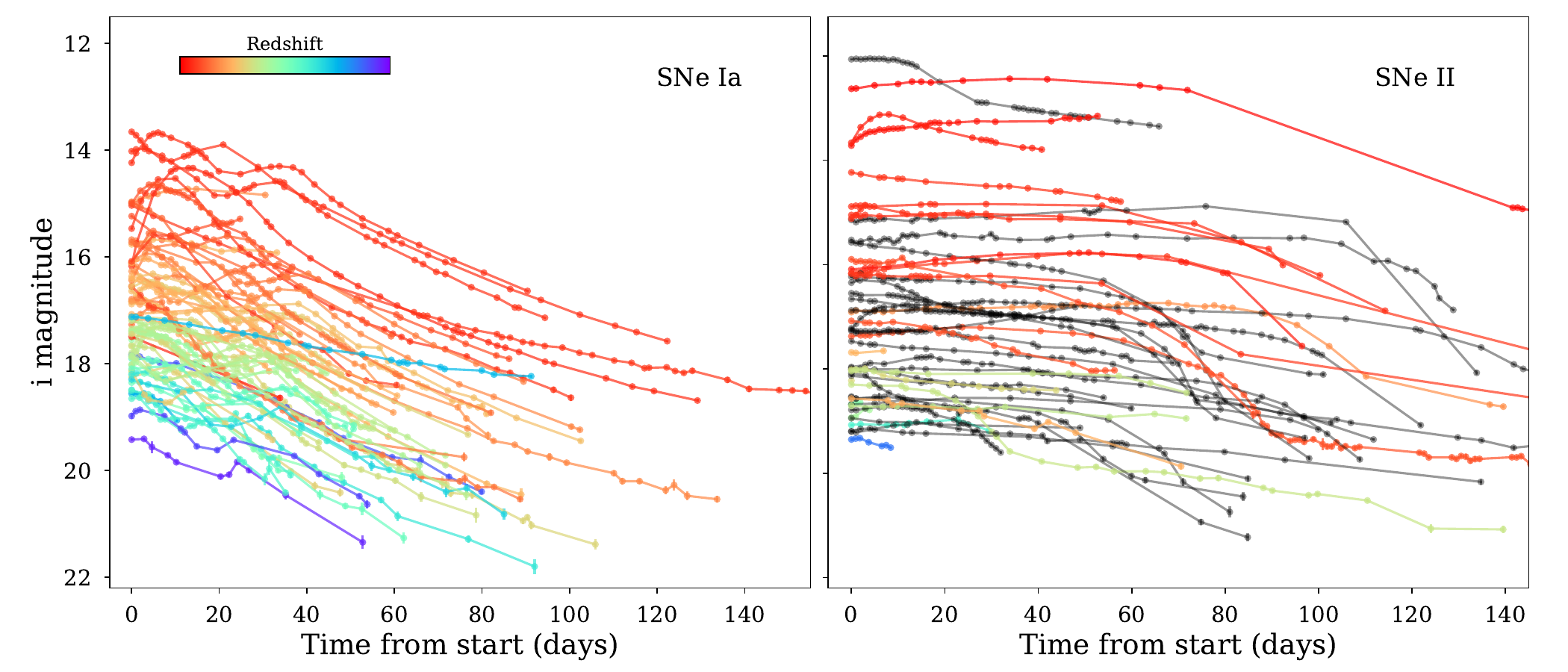}
\caption{
\emph{Left:} Light curves of 67 SNe Ia (thermonuclear supernovae), from the Carnegie Supernova Project.
Curves are colored by the redshift of the SN host galaxy according to the colorbar (with redshift increasing from left to right, spanning $z=0$ to $z=0.0835$).
The abscissa shows time since detection, as measured in the SN rest frame.
\emph{Right:} Light curves of 50 SN~II (core-collapse supernovae).
Light curves of 25 SNe with known redshifts are colored as in the left panel.
Gray light curves are for SNe with unknown redshifts (for these, the time axis is observer-frame time since detection).
\label{fig:CSPSN}}
\end{adjustwidth}
\end{figure}

\cfig{fig:CSPSN} shows 67 \snia\ light curves and 50 \snii\ light curves based on $i$-band photometry (a near-infrared passband), obtained by the Carnegie Supernova Project \cite{CSP17-SNIaPhotometry,CSP24-SNIIPhotometry} (these are subsets of the full samples of 134 \sneia\ and 94 \sneii, for clarity).
The light curves are colored according to the redshift of the supernovae host galaxies, serving here as a proxy for distance (all of the \sneia\ have host redshift measurements; only half of the \sneii\ sample has them).
The shapes of the light curves for these types of supernovae differ.
For example, for \sneia\ captured early enough, the $i$-band light curves are typically double-humped.
We would like to quantitatively describe the light curve distributions to facilitate \emph{light curve classification}.
While the full light curves shown here are fairly readily distinguished, these data span many months; astronomers need to classify light curves as well as possible as quickly as possible, e.g., to guide decisions on making expensive spectroscopic observations as early as possible after discovery.

\unskip

Careful examination of the light curves in the left panel reveals hints of important structure in \sneia\ light curves: when focusing on sets of light curves with similar redshifts (similar colors in the plot, corresponding to objects at similar distances), broader light curves (with shallower decays after the peak) tend to lie above light curves with faster decays (``above'' here corresponds to smaller magnitudes, i.e., brighter sources).
This ``broader-brighter'' relation can be quantified and exploited to estimate the luminosity of a \snia\ from the shape of its light curve.
We describe a Bayesian FDA treatment of this problem in \S~\ref{sec:SNIa}.

Some classes of variable astrophysical objects exhibit \emph{stochastic variability} rather than the periodic and transient variability of the examples just described.
An example is QSOs/quasars, the class of active-nucleus dominated galaxies mentioned above.
The stochastic variability is attributed to variability in the flow of matter from a turbulent accretion disk feeding a galaxy's central supermassive black hole.
Occasionally, some QSOs also exhibit transient emission, sometimes attributed to flares (sudden, large enhancements in accretion rate), and occasionally attributed to a tidal disruption event (TDE)---the destruction of a star by the black hole.

\begin{figure}[t]  
\begin{adjustwidth}{-.5\extralength}{0cm}
\includegraphics[width=14 cm]{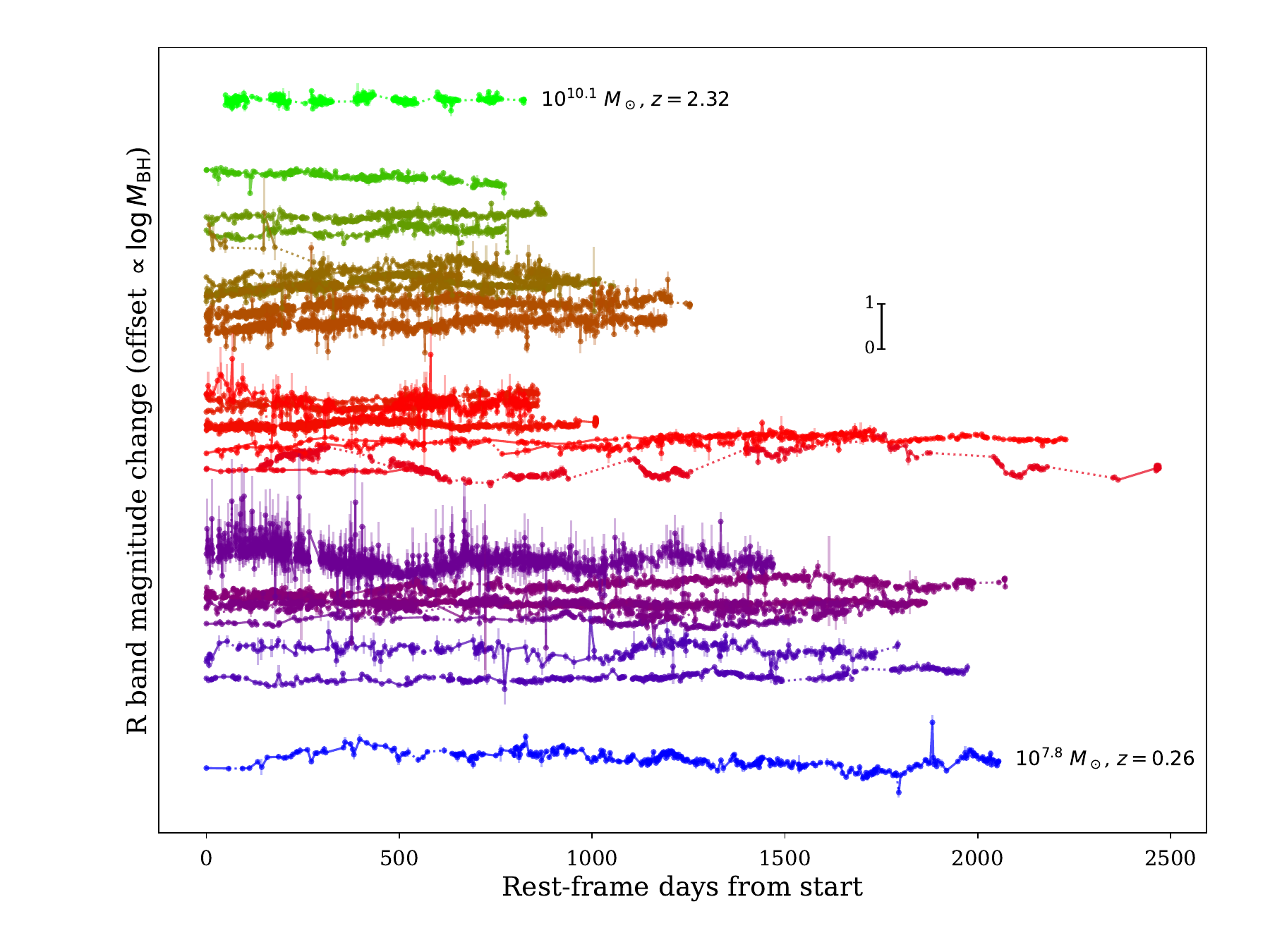}
\caption{
Light curves for 27 quasars observed in the field of the LMC by the MACHO project, shown as functions of time in the quasar rest frame.
Dots (with 1$\sigma$ error bars, many smaller than the dot marker) indicate irregularly-sampled $R$-band magnitudes and uncertainties; dashed lines connect observations with long segments of missing data.
Light curves are colored corresponding to estimated central black hole mass (estimated spectroscopically), using a blue-red-green color map.
To visually separate the light curves, the $R$-band magnitudes are offset proportional to black hole mass.
The vertical bar (center-right), spanning 0 to 1, shows a representative magnitude ordinate span for a single light curve.
The light curves with the smallest (bottom, blue) and largest (top, bright green) values of black hole mass are annotated with the values of the black hole mass value and quasar redshift, $z$.
\label{fig:MACHOQuasars}}
\end{adjustwidth}
\end{figure}

\cfig{fig:MACHOQuasars} shows light curves of 27 quasars observed in the field of the LMC by the Massive Compact Halo Object (MACHO) project (which surveyed the LMC for evidence of gravitational microlensing of LMC stars by dark massive objects passing along the line of sight) \cite{G+03-MACHOQuasars}.
The light curves are colored corresponding to estimated central black hole mass (estimated spectroscopically).
They are shown as a function of time in the quasar rest frame (where time passes a factor of $1+z$ more quickly than in the observer frame, for redshift $z$).

\unskip


These light curves visually resemble sample paths from simple stochastic processes, such as autoregressive models (albeit not uniformly sampled).
In fact, Kelly et al.~\cite{KSS11-AGN-DRW} showed that these light curves are well-modeled by a \emph{damped random walk} (DRW) model (equivalently, a continuous autoregressive model of order 1, CAR(1), or an Ornstein-Uhlenbeck process).
Individually fitting DRW models to such light curves shows that the DRW model parameters (the damping time scale and marginal variance) are correlated with black hole mass and luminosity (when those galaxy properties can be measured).
A small number of AGN (including quasars, the most luminous type) have been observed with much higher cadence than those shown here; those data suggest that the DRW model may not be appropriate for all AGN, with a more complex model required to account for short-time behavior \cite{KVR15-KeplerAGN-DRW}.
These studies model light curves \emph{individually}.
From an FDA perspective, collections of light curves (perhaps with similar covariate values, such as black hole mass estimates) could be modeled as independent sample paths from the same stochastic process, viewed as population distribution for light curves.
This could improve capability to detect structure beyond that of the DRW model, and improve classification of light curves (e.g., to distinguish quasars from other sources), particularly with short datasets.

\subsection{Beyond independent univariate function populations}

In the examples above, the function ensembles comprised \emph{univariate} functions that were exchangeable, i.e., able to be modeled as \emph{conditionally independent} draws from a population distribution.
Astronomers also observe function ensembles that are more complex than such collections of independent curves in two ways:
\begin{itemize}
\item \emph{Dependent function ensembles:} The vast majority of time-variable asrophysical objects have spectra that vary in time--\emph{dynamic spectra} or \emph{spectral time series}.
Observations of such objects produce collections of spectra that are ordered in time, evolving systematically (if still stochastically in some cases).
The collection of spectra across time from one such object should not be modeled as independent draws from a spectrum population; modeling should account for  dependence between spectra near in time (e.g., smooth spectral evolution).
\item \emph{Multivariate function populations:} Some modes of observation produce collections of related two- or three-dimensional functions, such as images of galaxies across a population, dynamic spectra for a specific source population (viewed as 2-D functions of wavelength and time), or 3-D ``data cubes'' or ``image cubes'' for a population of galaxies (i.e., spatially resolved spectra for each galaxy, functions of wavelength and two spatial or directional coordinates). 
\end{itemize}
Analysis and modeling of \emph{stellar activity}---particularly in the context of detection and characterization of exoplanets around distant stars---provides good examples of both forms of functional data complexity.

The varied phenomena comprising stellar activity---sunspots (dark regions of magnet\-ically-suppressed convection), prominences (long-lived arcs of magnetically-confined hot plasma), flares (explosive releases of plasma lasting minutes to hours), coronal mass ejections (abrupt eruptions of large amounts of plasma), and faculae and plages (bright filaments and patches in the lower and upper stellar atmosphere, resp.)---cause both the brightness of a star and the shape of its spectrum to vary in time.
The best-studied example is, of course, the Sun, though stellar activity is observed in less detail from many other stars.

\begin{figure}[ht!]  
\begin{adjustwidth}{-.1\extralength}{0cm}
\centerline{\includegraphics[width=17 cm]{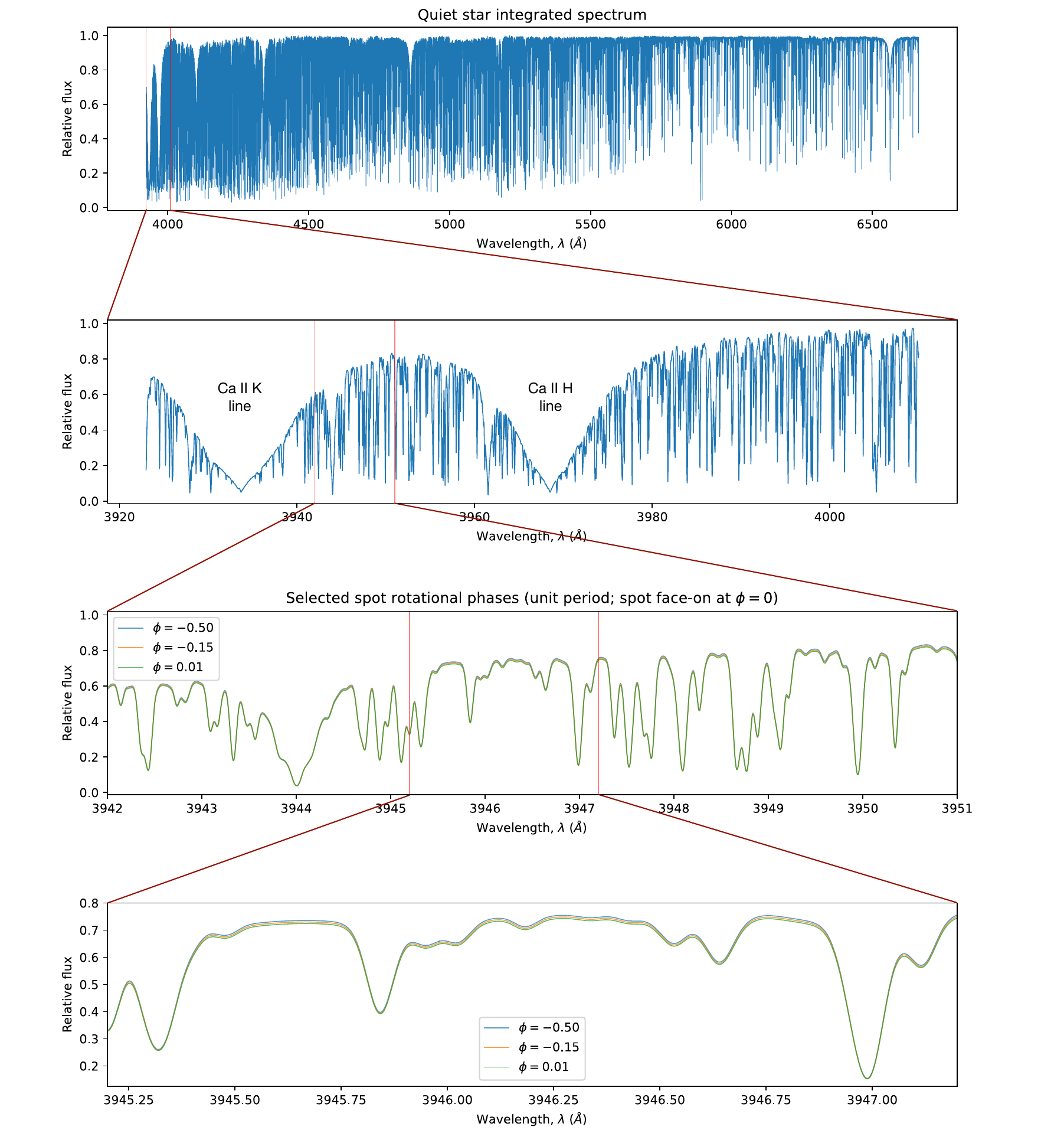}}
\caption{
Time-evolving spectrum of a Sun-like star with a single equatorial starspot occupying 1\% of the stellar surface, generated using SOAPv4.
Top two plots show the quite-star spectrum (i.e., with the spot hidden), over the full optical range (top panel), and zoomed in to the vicinity of the Ca~II H and K lines (2nd from top).
The bottom plots each show three spectra, at further zoom levels, with the spot at three rotational phases, $\phi$, as indicated.
\label{fig:SOAPSpecZoom}}
\end{adjustwidth}
\end{figure}   
\unskip

\cfig{fig:SOAPSpecZoom} shows simulated high-resolution spectra measured at different times as a large sunspot rotates across the face of a Sun-like star.
They were generated by the Spot Oscillation And Planet (SOAPv4) code, which uses actual measurements of solar spectra from quiet and active regions of the Sun to construct realistic simulated spectra reflecting the impact of sunspots and faculae with user-specified geometry \cite{C+25-SOAPv4}.
The simulation used here is for a large equatorial spot occupying 1\% of the stellar surface (the code can optionally include a spectral Doppler shift due to a companion planet; this capability is not used here).
The top panel shows the quiet-star spectrum (with the spot on the back hemisphere, thus out of sight), for reference.
Thousands of line features are visible, due to absorption of light by ionized gas in the solar atmosphere, with each line associated with different ionization states of different atoms and molecules.
The plot shows \emph{relative} (``normalized'') flux, found by dividing the spectrum by a smoothed version that interpolates over lines, to highlight the line features.
The next panel zooms in to the vicinity of the broad Ca~II (single-ionized calcium) H and K lines (corresponding to different electron transitions).
The bottom two panels each show three spectra at two further zoom levels. 
The three spectra have the spot at three rotational phases, $\phi$ (fractional rotation in $[-0.5, 0.5]$)---spot hidden at phase  $\phi=-0.5$ (as in the top two panels), spot face-on at $\phi=0$, and an intermediate case with $\phi=-0.15$.
The three spectra only become distinguishable at the highest zoom level:
the bottom panel reveals a barely discernable reduction in the
spectrum amplitude with the spot present, due to the spot's reduced temperature (compared to that of the quiet photosphere).


A collection of stellar spectra observed at different times comprises a \emph{functional time series}---a time-ordered series of related functions, here normalized high-resolution stellar spectra.
The spectra should not be viewed as independent sample paths drawn from a common underlying stochastic process in wavelength.
Rather, they may be viewed as a temporally correlated collection of functions of wavelength.
Modeling a time-ordered collection of functions like this is the domain of functional time series analysis, an example of what statisticians refer to as \emph{second-generation FDA} (with first-generation FDA restricted to scenarios with independently sampled one-dimensional curves).

\cfig{fig:SOAPSpecZoom} shows that the effects of modest stellar activity on stellar spectra are small for Sun-like stars.
But it is scientifically important to characterize these effects as accurately and precisely as possible, both for the insight they provide for the physical origins of stellar activity (which impacts Earth's space weather), and for the crucial impact they have on detection and characterization of planets around other stars (exoplanets).
Exoplanet masses are most precisely measured by detecting the reflex motion of the planet's host star as the star and planet jointly orbit the planetary system's center of mass.
The reflex motion is measured by detecting Doppler shifts in the locations of stellar absorption lines, due to the component of the motion along the line of sight (radial velocity, or RV).
For Earth-like planets orbiting Sun-like stars, the shifts are minute, corresponding to only a fraction of a percent of the width of a spectral pixel (there are tens of thousands of wavelength pixels in the spectra shown in \cfig{fig:SOAPSpecZoom}).
The shifts are only measurable by making extreme-precision spectral measurements and pooling data across hundreds or thousands of lines.
The small spectral changes induced by stellar activity are not small compared to the sources of uncertainty in such \emph{extreme-precision radial velocity} (EPRV) measurements.
In fact, our ability to model stellar activity is currently the limiting factor hindering the detection and characterization of Earth-like planets around Sun-like stars \cite{F+16-EPRV}.

\begin{figure}[t]
\begin{adjustwidth}{-.5\extralength}{0cm}
\centerline{\includegraphics[width=15 cm]{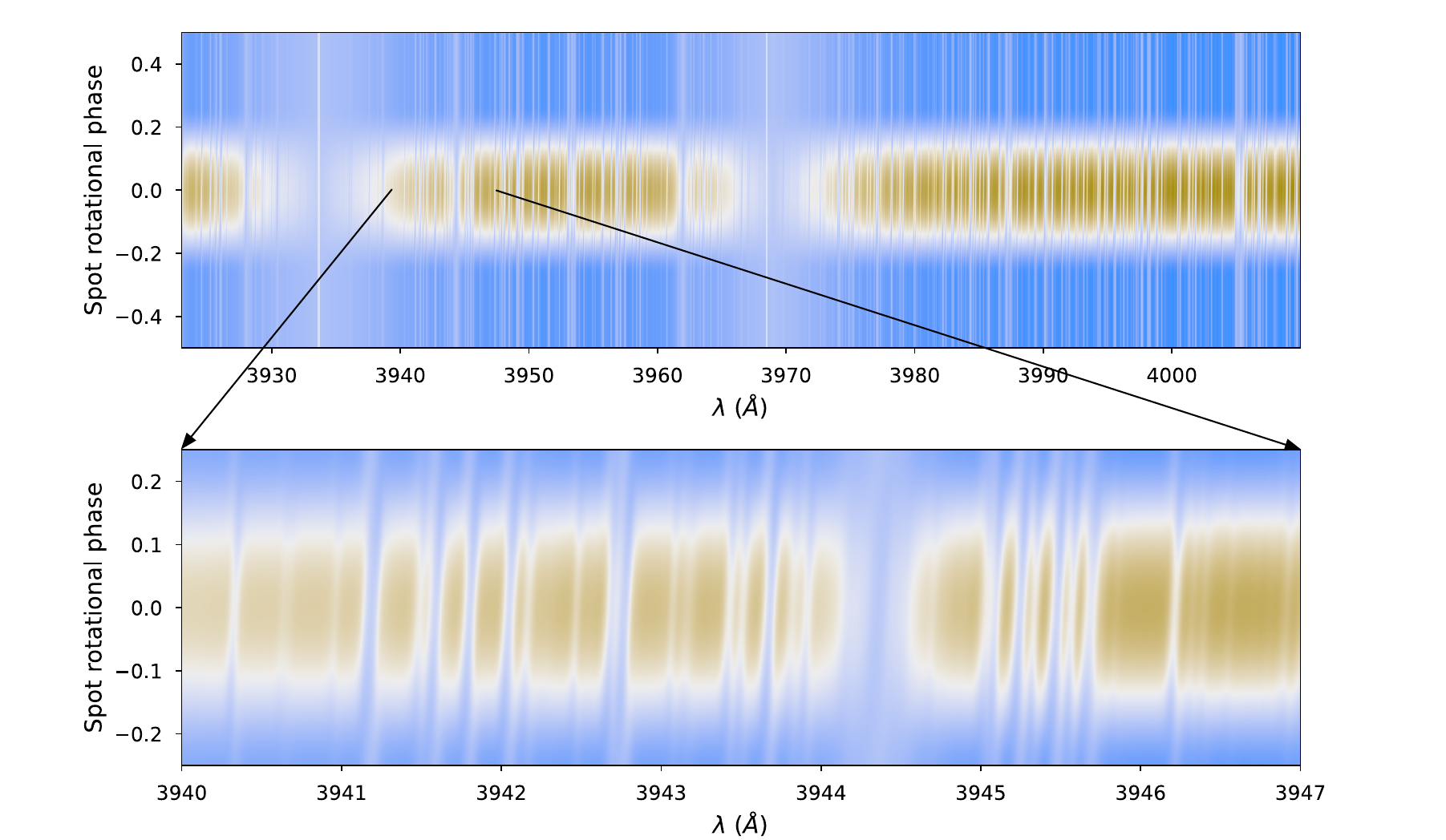}}
\caption{
SOAPv4 spectra as in \cfig{fig:SOAPSpecZoom}, but here for 100 phases, displayed as difference images. 
Time (phase) is vertical, and wavelength horizontal.
The color map uses brown hues (``ground'') for negative values, white hues for values near 0, and blue hues for positive values (``sky'').
The bottom panel zooms in on the indicated region in the top panel.
 \label{fig:SOAPSpecDiffImage}}
\end{adjustwidth}
\end{figure}

An important question is how similar the Sun's activity is to that of other, similar stars---so-called G dwarf stars.
To address this question, an alternative perspective is helpful, viewing a star's dynamic spectrum as a single 2-D function, and studying the distribution of such 2-D functions across a stellar population.
\cfig{fig:SOAPSpecDiffImage} shows the same evolving spectrum that was sampled at three phases in the previous figure, but now computed at 100 evenly-space phases in the interval $[-0.25, 0.25]$, corresponding to the spot just becoming visible, then face-on, and then rotating out of sight.
The spectra are displayed, not as a collection of curves, but as a 2-D false-color image, portraying a 2-D function (time increases from bottom to top).
To visualize the small changes in the spectrum, \cfig{fig:SOAPSpecDiffImage} specifically displays a difference image, the difference between the instantaneous spectrum and the time-averaged spectrum, for the wavelength regions covered in the two middle panels of \cfig{fig:SOAPSpecZoom}.

\unskip

This difference image view of the dynamic spectrum reveals more complexity that was apparent from viewing a few stacked spectra in \cfig{fig:SOAPSpecZoom}.
Different regions of the spectrum are affected by the spot in different ways, even in opposite senses. 
Particularly notable are the slanted streaks evident at high zoom level. 
These correspond to small, systematic shifts in line locations with time---spectral distortions that resemble a Doppler shift, with the potential to corrupt stellar RV measurements in exoplanet searches.
These distortions largely come from the spot slightly masking blue-shifted stellar emission as the spot emerges over the advancing limb (where the stellar surface is moving toward the observer), and later masking red-shifted emission as it moves to the receding hemisphere.

A population-wide study of stellar activity would jointly analyze multiple dynamic spectrum measurements like this---the type of second-generation FDA that treats multivariate rather than univariate functions.
The Sun's spectrum is easy to measure with high precision at high temporal cadence.
Extreme-precision spectroscopy of other stars is challenging, and can be done with only relatively sparse and irregular time sampling, presenting challenges for second-generation functional data analysis of stellar activity (see \S~\ref{subsxn:fda} for an example model architecture aiming to address this).

In our overview of light curve ensembles in astronomy, above, we presented ensembles of magnitude measurements in a single spectral passband.
In fact, the majority of light curve data in astronomy typically comprise measurements in multiple passbands, almost necessarily measured asynchronously when the same telescope is used for all measurements (since measurements in multiple passbands requires observing with different filters in the optical path).
The flux measurements for an object at a set of times and in a set of passbands, are measurements of \emph{linear functionals} of an underlying time-evolving spectrum (such flux measurements may be reported directly, or on the logarithmic magnitude scale used in several plots above; see footnote~\ref{fn:mag}).
Denoting the transmission function for band $\beta$ by $T_\beta(\lambda)$, the energy flux at time $t_i$ in band $\beta_i$ may be written,
\begin{equation}
F_{i,\beta_i} = \int d\lambda\; T_{\beta_i}(\lambda)\, f(\lambda,t_i),
\label{multiband-mags}
\end{equation}
where $f(\lambda,t)$ denotes the underlying evolving spectral flux density.
So, formally, modeling most light curve data requires implicitly or explicitly modeling an underly dynamic spectrum.
In this sense, astronomical light curve ensembles are in fact a type of second-generation functional data---a collection of functional (spectral) time series, albeit with the spectral dimension somewhat crudely measured.

\subsection{Beyond optical light curve and spectrum ensembles}

This ends our survey of the emergence of functional data in astronomy.
As lengthy as this section is, it only scratches the surface in regard to examples of functional data in astronomy.
This synopsis has focused on observations arising in optical astronomy.
Diverse types of functional data arise in other areas of astronomy, e.g., in studies of transient phenomena in radio astronomy (fast radio bursts) and high-energy astrophysics (X-ray bursts and gamma-ray bursts), quasi-periodic spectro-temporal variability in X-ray astronomy (quasi-periodic oscillators), and periodic variability with cycle-to-cycle diversity across the electromagnetic spectrum (radio and X-ray emission from pulsars).
Also, we have not here explored image ensembles, e.g., for studying the morphological diversity of galaxies, and including spectrally-resolved imagery, which arises in many areas of astronomy and is especially prevalent in submillimeter astronomy (producing spectral image cubes or hyperspectral data cubes).

\section{Functional data analysis}
\label{sec:FDA}

FDA seeks to discover and exploit patterns in the measured shapes of functions across an ensemble in a manner that accounts for diversity across the ensemble.
Important motivating problem areas in astronomy include:
\begin{enumerate}
\item Characterizing the diversity of light curves for a population of variable or transient objects:
For example, calibrating or generalizing the Leavitt period-luminosity law for Cepheids \cite{BHR25-LeavittLegacy}, or the luminosity-time scale law for Type~Ia supernovae, accounting for patterns and diversity in light curve shapes (see \S~\ref{sec:SNIa}).
\item Characterizing the diversity of galaxy spectra, e.g., to enable estimation of galaxy redshifts---Doppler shifts from cosmological expansion that provide distance estimates---from multiband magnitude measurements that provide significant but crude constraints on galaxy spectra (photometric redshifts).
\item Classifying irregularly sampled and possibly sparse light curves from observations of variable and transient objects, based on characterizations of well-observed prototypical objects, and including cases with light curves observed asynchronously in multiple spectral passbands.
\item Discovering new types of objects amidst ensembles of spectral or temporal (light curve) data, such as new populations (clusters of similar functions) or anomalous objects (outliers with respect to known populations).
\end{enumerate}
These tasks bear similarities to tasks with multivariate (vector-valued) data that one might address with regression and/or density estimation (tasks 1 and 2, possibly with dimension reduction steps), or with statistical or machine learning (ML) methods for supervised classification or unsupervised clustering or outlier detection (tasks 3 and 4).

One way to map these tasks to existing multivariate statistical or ML methodology would be to replace the basic functional data with a fixed-dimension set of a priori specified \emph{features} derived from the basic data.
Functional data analysis instead aims to frame such problems fundamentally in terms of studying \emph{distributions of functions}, producing ``function-focused'' counterparts to regression, density estimation, classification, clustering, and outlier detection.
FDA methods and model architectures can end up focusing on implicitly defined features of the observed functions; this hints at potential points of contact with feature-based ML methods (particularly deep learning methods that have feature learning capability) that we will briefly speak to in the concluding section.


To set the stage for a discussion of FDA methods for astronomy, here we review key aspects of the nature of functional data, and themes appearing in FDA modeling of such data.

\subsection{Mathematical and statistical character of functional data}

Functional data comprise \emph{measurements of many related functions}, related in the sense of being associated with objects considered to comprise a statistical population or ensemble, or of coming from repeated observation of functions from a single source.
In contrast to the vector data that is the subject of multivariate statistics, functional data may not be of fixed size (dimension) across the measured population.
Further, the data from each member function typically have a \emph{topology} (a sense of proximity among measurements), reflecting that the argument of the measured functions is continuous (e.g., time, wavelength, or spatial coordinates).
In contrast, the components of vector data often have no natural notion of order or proximity (e.g., between magnitude and direction data in a tabular catalog).

Until recently, most of the FDA literature in statistics has focused on univariate function ensembles.
In astronomy, common examples of univariate functional data include measurements of light curves (functions of time, $f(t)$), or of spectra (functions of wavelength, $f(\lambda)$), across populations.
Functional data can pertain to multivariate functions; such cases are sometimes labeled ``next-generation'' or ``second-generation'' FDA problems \cite{KS23-FDA2}, or as object-oriented data analysis (OODA) problems \cite{MD22-OODA-book}.
In astronomy, two-dimensional functional data arise in observations of dynamic (time-dependent) spectra, $f(\lambda,t)$, or of images of a class of extended objects, with $f(x,y)$ denoting image intensity as a function of spatial or angular coordinates.
Spatially-resolved spectroscopy produces ``image cube'' data---measurements of intensity as a function of both space, and wavelength, $f(x,y,\lambda)$.
Joint modeling of image cubes from a population of sources would be an even more complex example of second-generation FDA. 


In most cases, functional data are modeled as measurements of an unknown function at a set of specified sample points, e.g., light curves measured at a set of epochs, or spectra measured at a set of wavelengths corresponding to spectroscopic pixels.
But in some cases functional data comprise measurements of \emph{functionals} of an underlying function, in the sense of functional analysis (with linear functionals represented as inner product or convolution integrals).
A prevalent example in astronomy is photometric data obtained using multiple spectral bandpasses (``filters''); the flux expected in a band is the integral of the product of the source spectrum and the system transmission function for that band, shown above in \ceqn{multiband-mags}.
The double meaning of ``functional'' in such contexts is unfortunate but unavoidable---one could call such data ``doubly-functional.''


Functional data arise in diverse sampling regimes:
\begin{itemize}
\item  \emph{Sampling density:}
The functions may be densely or sparsely sampled.
Early FDA literature treated densely sampled data, a protypical example being a collection of handwritten digits or letters, where the data are so densely sampled that they could almost be considered to be curves rather than discete sets of measurements.
In astronomy, spectra are often densely sampled, but light curves (time series) are often sparsely sampled.
Dynamic spectra may be densely sampled in wavelength but sparsely sampled in time.
\item \emph{Sampling pattern:}
Across the ensemble, the measurements may be on uniformly-spaced, aligned grids, or they may be at irregular, unaligned, asynchronous sample locations, with different numbers of measurements for each member (typical in astronomy).
\item \emph{Sample completeness:}
The sampling may be complete across the ensemble (when there are shared grids), or there could be gaps or missing data, with missingness varying across the members.
The latter case is common in astronomy.
\end{itemize}

\noindent
Functional data are obtained in different measurement error regimes:
\begin{itemize}
\item Some functional data may be considered to be effectively noiseless; this was the case for the early prototypical FDA problems mentioned above.
In such settings, the statistical aspects of the problem arise solely from the statistical description of the ensemble of functions.
\item  Functional data often have measurement error.
Most commonly in the FDA literature, noise is modeled as additive, Gaussian, and homoskedastic (with the same standard deviation for all measurements), though possibly with unknown standard deviation.
In astronomy, measurement error is almost always heteroskedastic, though often with known standard deviations.
Measurement errors are often Gaussian to a good approximation, but sometimes are
more complex (e.g., reflecting Poisson counting uncertainties from photon counting data).
\end{itemize}

\begin{figure}[t]  
\begin{adjustwidth}{-.5\extralength}{0cm}
\includegraphics[width=16.25 cm]{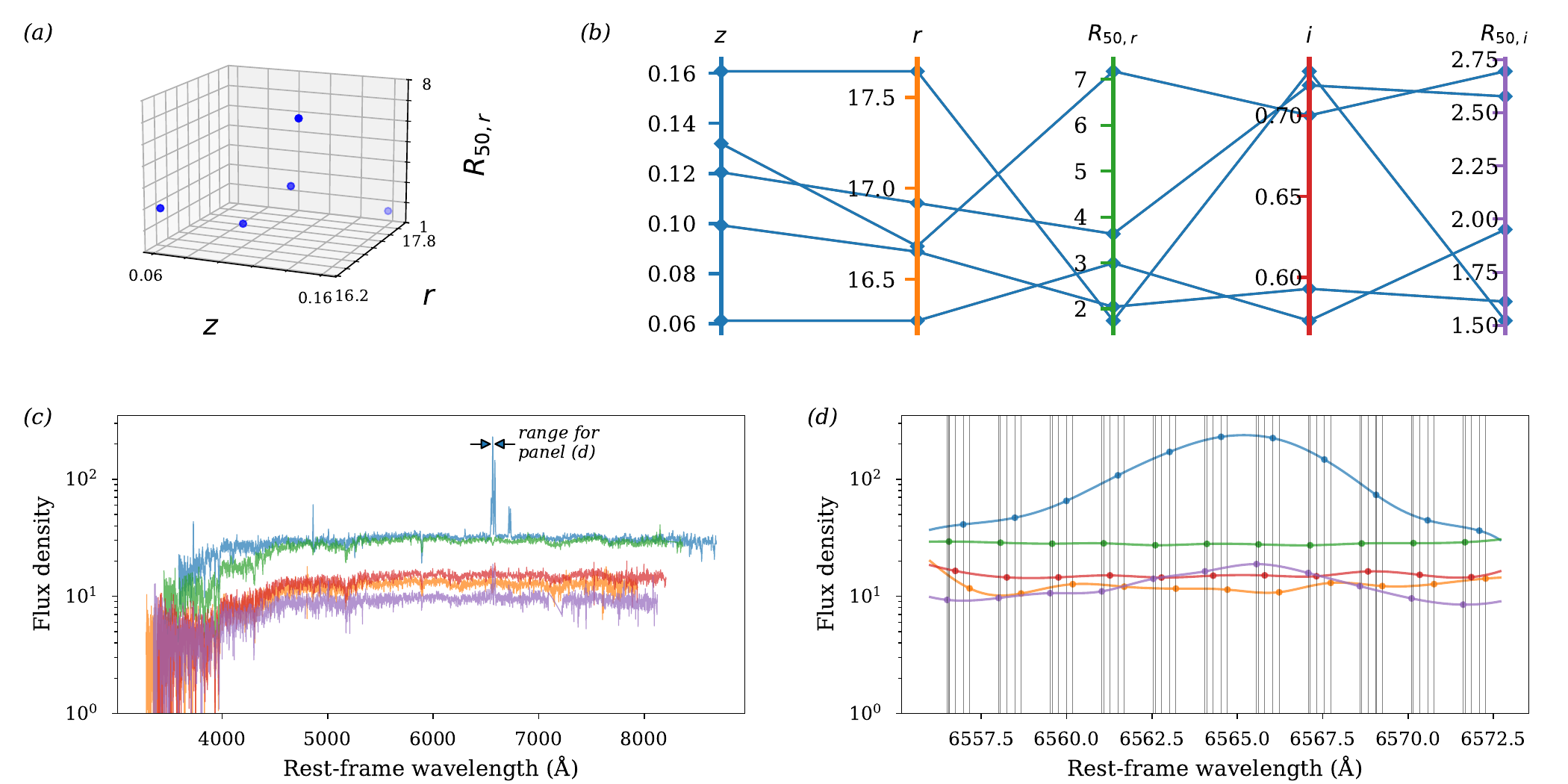}
\caption{
Visual illustration of similarities and differences between multivariate and functional data, using galaxy data from the SDSS Main Galaxy Sample (MGS).
\emph{Top row} shows multivariate data comprising five quantities measured for five galaxies: redshift $z$, $r$-band magnitude, $r-band$ galaxy size $R_{50,r}$, $i$-band magnitude, and $i$-band galaxy size $R_{50,i}$.
Each galaxy's properties define a point in a 5-D space.
\emph{(a):}~3-D orthogonal coordinate system (in projection) showing scatterplot of three of the five properties.
\emph{(b):}~5-D plot using parallel coordinates, with points represented as polylines.
Axes (all vertical) are drawn in different colors to emphasize they represent different types of quantities (e.g., with different units).
\emph{Bottom row} shows measured spectra for the same five galaxies, as a function of rest-frame wavelength.
\emph{(c):}~Spectra over the full measured range (colors distinguish the five sets of measurements).
Arrows show the wavelength range for the zoomed-in plot.
\emph{(d):}~Zoomed-in view of the spectra in \emph{(c)}, in the vicinity of a line feature.
Points show measured flux densities; curves show a cubic B-spline interpolant.
Vertical gray lines are ordinate axes at the sample locations for the various spectra; the sample locations are not aligned across the spectra
 \label{fig:MV+FuncData}}
\end{adjustwidth}
\end{figure}

Figure~\ref{fig:MV+FuncData} illustrates some of the basic aspects of functional data described above, to highlight similarities and differences between multivariate and functional data.
It makes use of \emph{parallel coordinates} (PCs) for visualizing high-dimensional data \cite{I12-HiD-ParallelCoord,W90-ParallelCoord}.
Two-dimensional scatterplots using conventional orthogonal coordinate axes are limited to depicting data with only two dimensions, or three dimensions via projection.
Plotting the coordinates of points along adjacent parallel axes produces a generalized kind of scatterplot that can display high-dimensional data in two dimensions.
A complicating consequence of the generalization is that a point in orthogonal coordinates appears as a \emph{polyline} in PCs (line segments connecting a points various coordinates).


\unskip

The two panels in the top row of Figure~\ref{fig:MV+FuncData} show scatterplots of five-dimensional multivariate data providing measurements of the properties of five galaxies from the SDSS MGS.
The left panel depicts a subset of three quantities as points in a three-dimensional orthogonal coordinate space (each defining a vector from the origin).
The right panel depicts all of the data as polylines in a five-dimensional PC space.
Note that the axes show different kinds of physical quantities (reflected as different colors for the axes); each axis has different units, and the order and proximity of the axes is arbitrary.

The two panels in the bottom row show functional data for the same five galaxies---their SDSS spectra.
The left panel shows a conventional plot of spectra as sampled functions (flux density vs.\ wavelength).
The right panel zooms in to show that such a plot can be understood as a PC plot, with each measurement assigned its own vertical coordinate axis (the ordinate for the function at the measured sample locations).
Here, the vertical axes for each measurement are shown in the same color (gray), indicating each axis is a copy of the same space (flux density); this is in contrast to multivariate data, where the spaces for each coordinate of vector data typically differ.%
\footnote{Note that a stochastic process is joint distribution on an arbitrary number of random variables in a \emph{common} probability space, labeled by values from an index set, e.g., time, space, wavelength, etc..
The PC depiction graphically displays the suitability of stochastic processes as modeling tools for functional data, as described below.}
Further, the horizontal separation between the axes is meaningful, reflecting the common continuous argument of the sampled functions (here, wavelength).
Finally, the right panel also shows how measurements may be misaligned for the different members of the functional data ensemble.

\subsection{Functional data analysis fundamentals}
\label{subsxn:fda}

Functional data \emph{analysis} refers to statistical methods tailored for explanatory and predictive modeling using functional data.
Typically there is significant diversity in the function shape across the population, requiring use of high-dimensional models.

The fundamental units or ``atoms'' being modeled in FDA are continuous functions rather than the fixed- and finite-length vectors of variables that are the subject of multivariate data analysis and spatial statistics.
But FDA builds on the methods and intuition of vector-based statistical modeling, exploiting ideas such as basis expansions to link continuous function modeling to working in an underlying (latent) vector space (possibly of infinite dimension).


Common inference and prediction tasks in multivariate statistics include estimating the joint distribution of components of the data vector (univariate or multivariate density estimation), finding relationships between the components (correlations, regression laws/scaling laws, and conditional distributions), and clustering and classification.
FDA replaces estimating the joint distribution of the data vector with learning the distribution for the function population (e.g., as a stochastic process or via some other architecture; see below)---one could call this \emph{functional density estimation}, though the notion of a probability density over functions is subtle \cite{DH10-PDFRandomFuncs}, so this terminology is not common.
For inference of relationships, \emph{functional regression} technology has been developed for a variety of regression tasks involving measured or predicted functions and scalar or vector covariates (predictors) or responses (predictands):
\begin{itemize}
\item Scalar- or vector-on-function regression aims to predict a finite-dimensional quantity of interest from a measured function; predicting a variable star's (time-averaged) luminosity from its light curve is an astronomical example.
\item Function-on-scalar or -vector regression views the functions as responses rather than predictors.
An example is predicting the spectrum of a galaxy from a finite set of features or coordinates (defined physically or in a data-driven manner); see \S~\ref{sec:SEDPop}.
\item Function-on-function regression uses one type of function to predict another, e.g., using a spectrum (perhaps time-averaged) to predict a time series.
\end{itemize}
Similarly, \emph{functional classification} and \emph{functional clustering} are the FDA counterparts to classification and clustering in multivariate statistics and supervised/unsupervised ML.



In the underlying theory motivating and validating many FDA methods, infinite dimensional \emph{Hilbert spaces} play an important role.%
\footnote{We are here restricting to so-called separable Hilbert spaces, which have countably infinite dimension.
Nonseparable Hilbert spaces, of higher cardinality, are not important for FDA applications.}
While the theory seldom plays an explicit role in applications, many astronomers have some familiarity with Hilbert space from exposure to quantum mechanics, and some insights from the theory are thus accessible.
A Hilbert space is a vector space---a space whose elements can be added, and multiplied by scalars---albeit one whose elements can be continuous functions, corresponding to vectors of infinite dimension.
A key property of a Hilbert space is the existence of an inner product between two functions (e.g., defined by an integral over their product, possibly with a weight function), mimicking the familiar Euclidean vector inner ``dot'' product.
The inner product provides a notion of mutual projection between functions in the space (with orthogonality corresponding to zero inner product), which enables quantification of similarity between functions.
The inner product of a function with itself gives the squared magnitude of the function, an analog to the length of a vector in Euclidean space.
All functions in a Hilbert space must have a finite magnitude (analogous to requiring that a Euclidean vector have no infinite coordinates).
Restricting to a function space with these properties (a subset of all possible functions) allows representation of a large family of functions via a superposition of a \emph{countable} number of orthogonal basis functions \cite{KS13-HilbertSpaceSigPro}.
The coefficients defining the superposition live in a vector space (of finite or countably infinite dimension).
Thus, working with superpositions of orthogonal (or orthogonalizable) functions provides a path for transferring intution and methodology from vector-based multivariate statistics to the modeling of functional data.

Key statistical modeling tools used in FDA include:
\begin{itemize}
\item Parametric and semiparametric expansions in a \emph{pre-specified basis}, e.g., splines, a Fourier basis (orthogonal sines and cosines), or wavelets.
\item \emph{Stochastic process models}, such as Gaussian processes, viewed as population distributions for functions, with the measured functions treated as independent sample paths from the process.
\item Expansions in a data-derived basis, typically found using \emph{functional principal components analysis} (FPCA).
Conventional PCA arises in multivariate statistics for finding basis vectors for vector-valued data, treated as independently and identically distributed (IID) samples from a multivariate joint density.
FPCA arises in FDA for deriving basis functions for a population of functions, treated as sample paths from a stochastic process written as a Karhunen-Loeve expansion (KLE, a random superposition of eigenfunctions of the process covariance function).
The resulting empirical KLE, when computed on a discrete grid, uses linear algebra resembling that of PCA, despite having a different rationale.
\item Dimension reduction and manifold learning, to discover low-dimensional descriptions of a function ensemble, using FPCA or techniques from machine learning.
\end{itemize}

In some cases (rare in astronomy), functional data measurements are available on a shared dense, aligned grid.
In such settings, in principle the finite set of function values measured on the grid may be used for modeling.
But the FDA perspective always views the underlying measured object as a function of a continuous argument.
For example, splines express a function in terms of function values on a grid (the spline nodes).
In FDA, the focus is on the continuous representation (e.g., the connected set of cubic polynomials for cubic splines), not on the discrete set of function values that happen to be convenient for expressing that function.
In particular, theory pertaining to the accuracy (bias) and variability (variance) of estimates uses the complete true and estimated functions (not just values at nodes) for comparing estimates to true functions.

\subsection{Common model architectures}
\label{subsxn:arch}

Statisticians have devised a wide variety of model architectures to capture various types of relationships involving function populations, including describing similarities and diversity among functions across a population (possibly accounting for different types of groupings), and relationships between function and scalar or vector properties associated with the objects or subjects comprising the population.
Here we briefly describe a few representative architectures, guided by the kinds of populations and associated scientific problems surveyd in \S~\ref{sec:Tab2Func}.

A common parametric or semiparametric FDA model architecture is a \emph{functional mixed effects model} (FMEM), where each function in the population is modeled as the sum of two components:
\begin{itemize}
\item A \emph{fixed effect}: A shared, template-like component whose shape is set by a parameter shared across the population (possibly also depending on observable covariates or predictors---scalar or vector properties of the object or subject producing the functional data---such as a measured period for time series data).
\item A \emph{random effect}: A peculiar (i.e., object-specific) component specified by latent (unobservable) parameter values unique to the object (or to its subgroup, in populations with subgroups), accounting for diversity across the population.
\end{itemize}
For example, for modeling light curves observed from a population of periodic variable stars, the period would be a covariate, the fixed effect would describe the part of the light curve shape common to objects with similar periods (a template of sorts), and the random effect would describe the diversity of light curve shapes with respect to the shared, template-like behavior (see below).

The fixed/random terminology originates from frequentist statistics, where only the parameters in the peculiar component are considered ``random variables'' (in the frequentist sense of varying statistically in repeated sampling).
In Bayesian settings, where ``random variable'' is interpreted more generally to mean ``uncertain quantity,'' the parameters of both components are random, and the \emph{shared} vs.\ \emph{peculiar} terminology seems more apt.

As an example, for function $F_o(x)$ associated with object $o$ ($o=1$ to $N$) observed at locations $x_{ol}$ ($l=1$ to $M_o$), a functional mixed effects model jointly modeling measurements $d_{ol}$ of the $N$ related functions could be written as follows:
\begin{equation}
\begin{split}
d_{ol} 
 &= F_o(x_{ol};\beta,\chi_o) + \eps_{ol}\\
 &= G(x_{ol};\beta) + g(x_{ol};\chi_o) + \eps_{ol}
\label{FDA-mixed}
\end{split}
\end{equation}
with measurement errors $\eps_{ol}$, shared parameter $\beta$, and peculiar parameters $\chi_o$.
Here $G(x_{ol};\beta)$ is the fixed effect, representating shared template-like behavior across the population, and $g(x_{ol};\chi_o)$ is the random effect, representing object-specific function behavior, with function diversity arising from the distribution of the (latent) $\chi_i$ parameters.
For \emph{parametric} FMEMs, these functions are typically represented as finite-dimension basis expansions, using either a pre-defined basis, or an empirical basis found via FPCA (the parameter vectors $\beta$ and $\chi_o$ are then the expansion coefficients).
For \emph{nonparametric} FMEMs, the functions may have formally infinite expansions, or the random effect functions may be specified as sample paths from a stochastic process (the latent $\chi_o$ parameters may then be implicit, corresponding to the random variables underlying the stochastic process).%
\footnote{Similarly, nonparametric FMEMs can be constructed using stochastic differential equations (SDEs), where the deterministic part of the SDE can provide a fixed effect component (presuming its free parameters are global), and the random part provides random effects; these are known as stochastic differential equation mixed-effects models (SDEMEMs).}


If object-specific covariates are measured---say, a period in a time series context, as arises in analyses of periodic variable star light curves---those covariates may be included in the fixed and/or random effect components, so that objects with the same covariate value (rather than all objects) share some functional behavior.%
\footnote{Strictly speaking, the period of a variable star is estimated from the light curve (e.g., via Fourier methods), so it is technically not a covariate.
But Hall, Reimann, and Rice \cite{HRR00-NonparamEstPerFunc} show that even when a periodic light curve's shape is uncertain, its period can be measured with surprising accuracy with even modest-length samples, and can be treated as a known quantity for the purpose of estimating the light curve shape.}



An alternative architecture is to consider the functions to be independent sample paths drawn from a common underlying stochastic process.
This architecture is particularly natural for settings like the quasar light curves of \cfig{fig:MACHOQuasars}, where there is no apparent shared template-like behavior, and one instead is aiming to find and characterize statistical similarity between apparently stochastic functions.
For a stochastic process, $\SP(\theta)$, with parameters $\theta$ (e.g., specifying a covariance function or other properties of the process), such an architecture could be expressed as
\begin{equation}
\begin{gathered}
d_{ol} 
 = F_o(x_{ol}) + \eps_{ol},\\
F_o(\cdot) \sim \SP(\theta).
\label{FDA-SP}
\end{gathered}
\end{equation}




These architectures have generalizations to second-generation scenarios, such as the modeling of a dynamic spectrum, $F(\lambda,t)$, as a time-ordered sequence of dependent (one-dimensional) random spectra, or modeling a population of (two-dimensional) dynamic spectra.
Modeling higher dimensional functions adds computational expense, motivating architectures that enforce sparsity in some way (ideally motivated by both mathematical considerations and knowledge of the underlying physics).
An example is adoption of a \emph{separable expansion}, e.g., writing a two-dimensional function as a sum of products of one-dimensional functions, such as
\begin{equation}
F(\lambda,t) = \sum_\beta c_\beta\, \psi_\beta(\lambda)\,\chi_\beta(t),
\label{HSD}
\end{equation}
where $\psi_\beta(\lambda)$ denotes normalized spectral basis functions (``speclets''), $\chi_\beta(t)$ denotes normalized temporal basis functions (``modulators''), and the $c_\beta$ are expansion coefficients.
The \emph{asymmetric Hilbert-Schmidt decomposition} (HSD) provides a mathematical motivation for such an expansion \cite{S93-HistorySVD,S11-FredholmHilbertSchmidt,TT13-Chebfun2D,TLR26-EPRV-HSD}.
Schmidt showed how to construct such an expansion for functions in a Hilbert space, with the speclets and modulators satisfying a set of mutual integral equations resembling eigenvalue equations; the analog to eigenvalues are called singular values.
The expansion will be infinite in general, but Schmidt showed that a finite truncation of the expansion, keeping terms associated with the largest singular values, is optimal in terms of integrated mean squared error for approximating $F(\lambda,t)$.
Solving a discretized version of the integral equations (using quadrature rules for the integrals) amounts to performing a matrix singular value decomposition (SVD).
This relates the HSD to principal components analysis (PCA) and FPCA, but with a quantitatively important distinction: discretization of the integrals (over $\lambda$ and $t$ in this example) introduces factors that account for spacing of the wavelength and time sample points, a feature absent from PCA.
This enables more accurate modeling of smoothly varying spectra from unevenly sampled data, using an empirical separable basis constructed via the HSD.

\begin{figure}[t!]  
\begin{adjustwidth}{-.1\extralength}{0cm}
\centerline{\includegraphics[width=.8\textwidth]{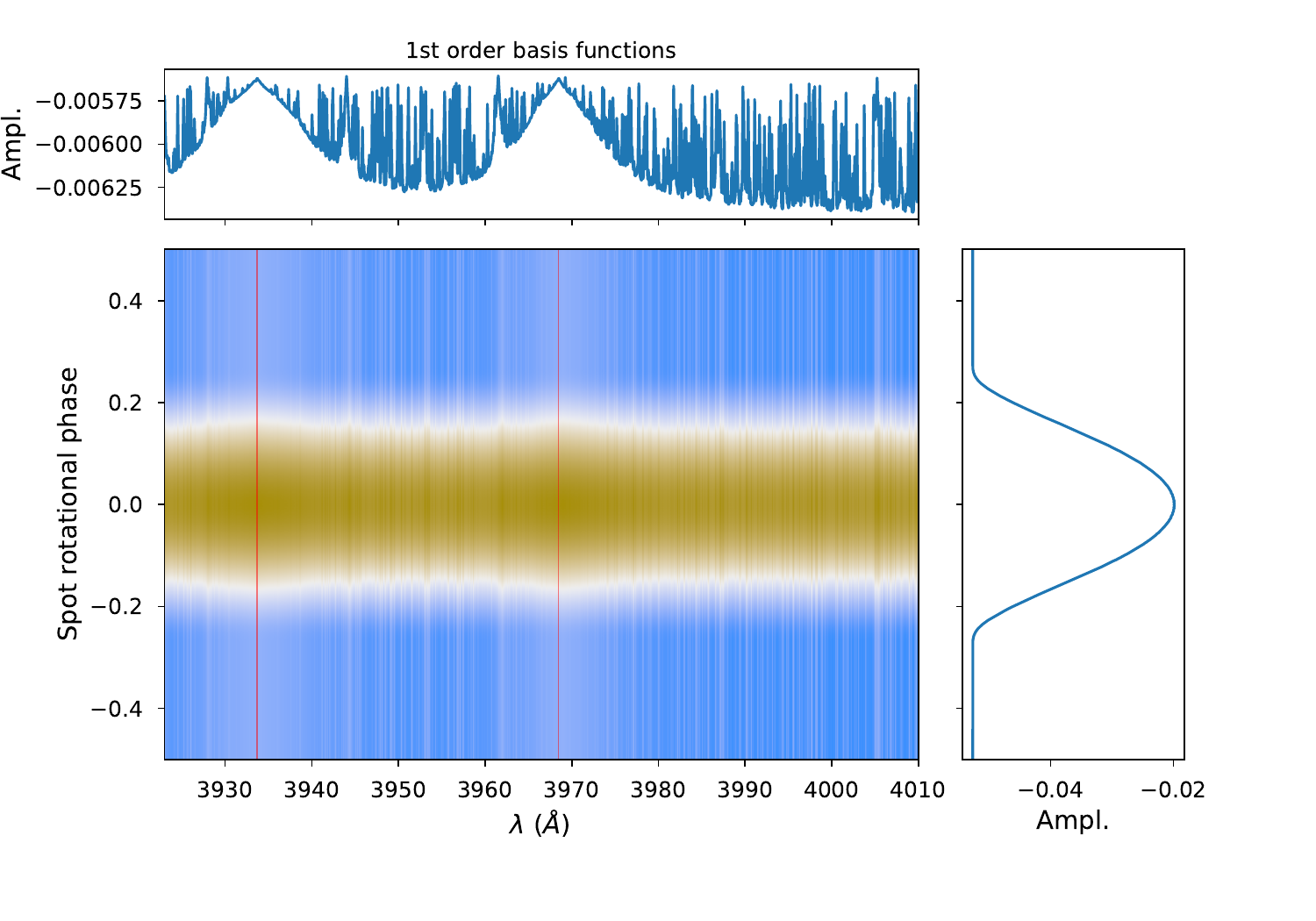}}
\centerline{\includegraphics[width=.8\textwidth]{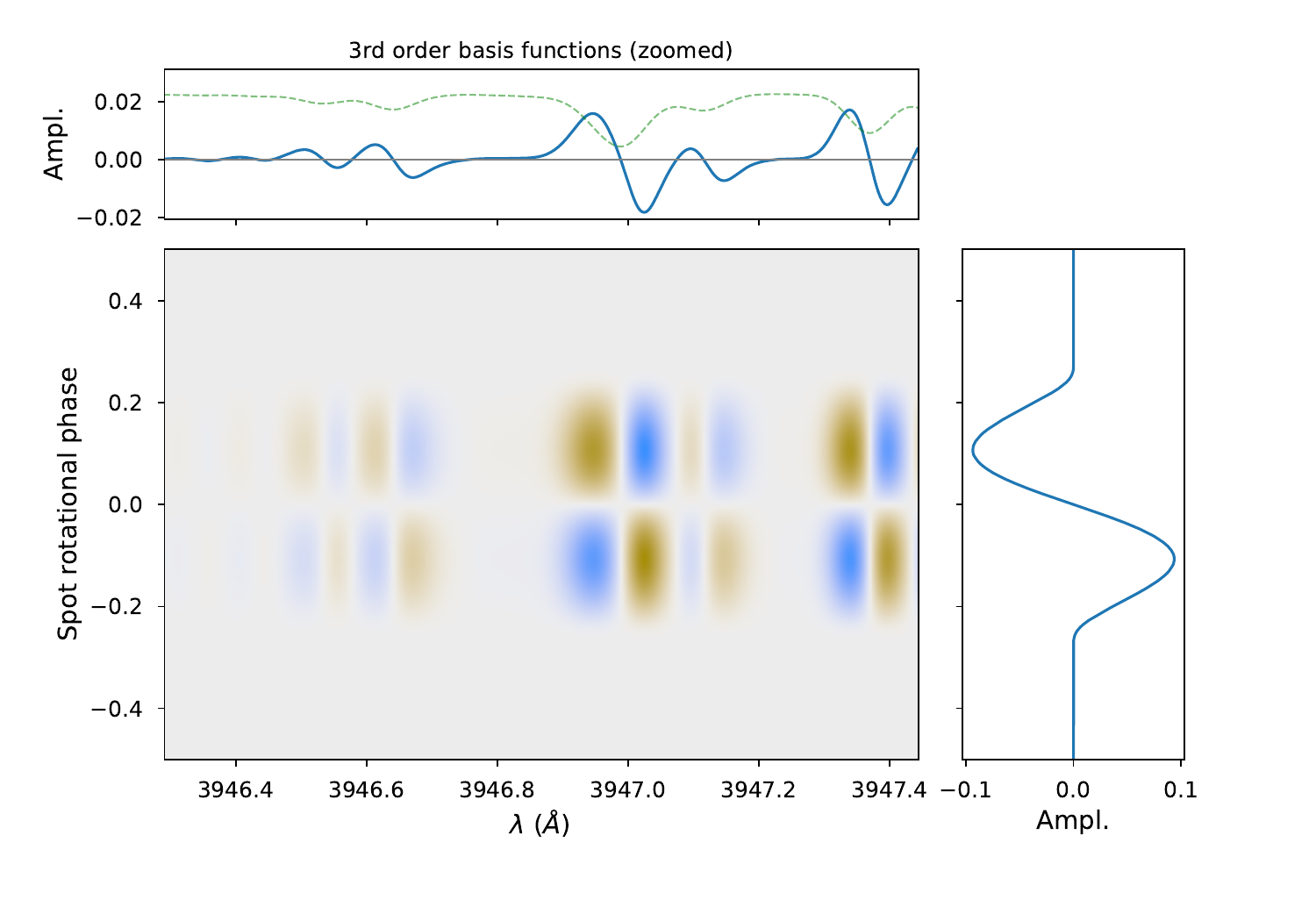}}
\caption{\small\baselineskip=11pt
Illustration of terms in the asymmetric Hilbert-Schmidt decomposition of a bivariate function, here the dynamic spectrum of a Sun-like star described in \S~\ref{sec:Tab2Func}, focusing on the Ca~II line region depicted in \cfig{fig:SOAPSpecZoom} (second panel from the top) and \cfig{fig:SOAPSpecDiffImage}.
Top 3-panel figure shows the leading term in the HSD.
The top panel shows the speclet, $\psi_1(\lambda)$; the right panel shows the modulator, $\chi_1(t)$ (as a function of rotational phase), and the middle panel shows their outer product, $\psi_1(\lambda)\chi_1(t)$, as a difference image.
The color map for the difference image uses brown hues (``ground'') for negative values, white hues for values near 0, and blue hues for positive values (``sky'').
Bottom 3-panel figure similarly shows the third-order term in the HSD, with the panels depicting $\psi_3(\lambda)$, $\chi_3(t)$, and $\psi_3(\lambda)\chi_3(t)$, zoomed in in wavelength to better reveal the structure in this term.}
\label{fig:HSD}
\end{adjustwidth}
\end{figure}

\cfig{fig:HSD} depicts how the HSD works for the dynamic spectrum of a Sun-like star exhibiting stellar variability from an equatorial sunspot, as escribed in \S~\ref{sec:Tab2Func} (figures \cfig{fig:SOAPSpecZoom} and \ref{fig:SOAPSpecDiffImage}), from a demonstration of the HSD described by Tao et al.\ \cite{TLR26-EPRV-HSD}.
The top 3-panel figure shows the leading term in the HSD.
It captures the overall small modulation of line strengths over a rotational period.
The bottom 3-panel figure (zoomed in in wavelength) similarly shows the third-order term in the HSD.
It exhibits a more interesting modulation (shown in the modulator plot in the right panel), changing sign as the spot moves toward (negative phases) and away (positive phases) from the observer.
The difference image panel shows that the net effect of this term is to cause an oscillating shift in the locations of many line features.
This kind of behavior can mimic the periodic Doppler shift that a planet can induce, so it is important to model it as accurately as possible.
The HSD provides a promising model architecture for this task.




\subsection{Bayesian functional data analysis}

Bayesian FDA frames FDA tasks within the Bayesian paradigm, where modeling of populations relies on hierarchical Bayesian (HB) methods.
Bayesian methods treat unknown parameters as well as data as ``random'' (in the sense of uncertain a priori), requiring introduction of prior distributions for parameters.
In HB population modeling, parameters that describe the population as a whole are dubbed \emph{hyperparameters}, and the prior distribution for such a parameter is sometimes called a hyperprior.
Object-specific parameters, not directly observable, are often called \emph{latent parameters}.
Whereas a hyperparameter has a single value in an HB model (globally controlling population properties), latent parameters are replicated across the population, so the prior distribution for the latent parameters is a population distribution, i.e., a statistical description of diversity across the population.
The population distribution hyperparameters may in fact be of key interest as descriptors of functional diversity.

In the mixed effects model of \ceqn{FDA-mixed}, $\beta$ is a hyperparameter controling the shared template function, and the $N$ values of $\chi_i$ are latent parameters, viewed as drawn from a population distribution that itself may have parameters $\theta$ (e.g., a population mean and standard deviation for the $\chi_i$ distribution).


Bayesian FDA may instead adopt the stochastic process modeling architecture, for example, specifying that the modeled functions are sample paths from a Gaussian process (GP) \cite{SCQ11-GPFDA}.
GP \emph{regression}---fitting measurements of a \emph{single} function with a GP---has become popular in astronomy, particularly in time series settings \cite{AF22-GPRegrnAstro}.
In a regression setting, where one has data from only a single sample path drawn from the GP, the covariance function controlling the flexibility of the GP must be highly constrained a priori (e.g., of squared exponential or Matern form, fully specified by two or three parameters).
Typically very simple covariance functions, with just two or three adjustable parameters, are used.
In contrast, in GP FDA, data from \emph{many} draws from the GP are analyzed \emph{jointly}.
This enables adopting much more flexible GP covariance structures, including nonstationary covariance functions with many parameters.
We provide an example below, in the context of modeling a population of galaxy spectra.

With either architecture, an important advantage of Bayesian FDA is how the statistical phenomena of \emph{shrinkage} and \emph{borrowing strength} arise naturally within a hierarchical Bayesian setup.
The joint, fully probabilistic analysis enables accumulation of information both along the paths of individual functions, and across the many paths comprising the function population \cite{MC06-WaveletFDA,M15-FDARegrn}.
The sharing of information across paths improves the accuracy with which each individual function can be recovered.
And the fully probabilistic framework allows inference at the population level (of hyperparameters, such as parameters defining scaling laws) to account for uncertainties in the individual function estimates more thoroughly than is possible if one were to naively fit or smooth each object's function data independently, and then use the independent best-fit estimates to characterize the population.
In fact, in many settings the latter approach is known to be statistically inconsistent, converging away from the truth as the number of objects measured increases.

This overview merely scratches the surface of FDA.
Good entry points with broad coverage of FDA include the influential early monographs of Ramsay and Silverman and their collaborators \cite{RS05-FDABook,RHG09-FDA-R-MATLAB}, an FDA handbook \cite{FR11-FDAHdbk}, and recent review articles (including coverage of second-generation FDA) \cite{M15-FDARegrn,WCM16-FDAReview,KS23-FDA2}.
Zhao, Marron, and Wells \cite{ZMW04-FDA-LDA} provide an accessible tutorial on basic FDA from the perspective of longitudinal data analysis (LDA, the analysis of data from repeated observations of a population of subjects over time), with simple illustrations of the use of FPCA and Fourier methods for structure discovery (they also briefly note the relevance of parallel coordinates to visualization of functional data).
The monograph by Hsing and Eubank \cite{HE15-TheoreticalFDAOperators} covers the theoretical foundations of FDA, with thorough technical treatments of the Hilbert space and stochastic process perspectives.
M\"uller \cite{M16-Hall+FDA} provides an illuminating overview of the history of FDA up to 2016.
For specifically Bayesian FDA methodology, only a few works are available with introductory content; see the review of Morris \cite{M15-FDARegrn} and the monograph on GP FDA by Shi et al.\ \cite{SCQ11-GPFDA}.
Besides covering formal statistical modeling, several of these works also cover practical FDA tasks such as function registration.



\section{FDA in astronomy}
\label{sec:AstroFDA}


As noted in the Introduction, much emerging astronomical data would be recognized as functional data by statisticians, especially large spectroscopic and light curve catalogs being produced by current and forthcoming surveys.
To date, only rarely have astronomers explicitly drawn on methods in the FDA literature to analyze such data.
Of course, in the statistics literature, FDA methodology was not suddenly invented; it arose in response to changes in data modalities, becoming identified as a new subdiscipline worthy of a new name several years after the foundational ideas began appearing.
Similarly, astronomers have developed what is essentially FDA technology in response to challenges posed by new functional datasets.
For example, Tanvir, Hendry, Kanbur and collaborators developed a technique for fitting the light curves of Cepheid periodic variable stars that is very similar to FPCA \cite{T+05-CepheidLCFit,K+02-CepheidPCA}.
Mandel and collaborators developed and continue to refine a model for the population of Type~Ia supernova transient light curves that implements Bayesian GP FDA for dynamic spectra (bivariate functions of time and wavelength) \cite{M+22-HBSED-SNIa}.
This work is a good example of second-generation FDA in astronomy---and its initial development predates the emergence of second-generation FDA as an established subdiscipline.
We provide an overview of this work in \S~\ref{sec:SNIa}.

The earliest explicit use of FDA in astronomy that we are aware of is a 2011 study of the light curves of long-period Mira variable stars, reported in the statistics literature by a collaboration led by statistician Woncheol Jang and astronomer Martin Hendry.
They used FPCA and functional clustering to identify Mira variables that have periods that change significantly in time \cite{P+11-MiraFDA}.
In work reported in 2014, a statistician-led team at Carnegie Mellon University collaborated with an astronomer to use FDA to predict the unobscured continua of quasar spectra to enable estimation of damping of the spectrum by intervening gas, which produces a broad, complex spectral absorption feature, the Lyman-alpha forest \cite{C+14-LyAlphaFDA}.
In a more recent collaboration between astronomers and statisticians, Patil et al.\ used FPCA to analyze nearly 19,000 stellar spectra from SDSS observations of an open star cluster, to quantify the complexity of spectral diversity \cite{P+22-StellarSpecFPCA}.
Ten functional principal components are needed to characterize the statistically signficant diversity, and the FPCA scores can be used to estimate physical parameters describing the chemical composition of stars in the cluster.



Bayesian FDA architectures are relatively new to astronomy.
In various teams, the authors have been developing Bayesian FDA methods in five astrophysical problem areas:
\begin{itemize}
\item Modeling the multiband light curves of Type~Ia supernovae using a Gaussian process model for the underlying dynamic spectra;
\item Modeling the periodic light curves of Cepheid variable stars using a Fourier mixed-effects model;
\item Classification of variable star and transient object light curves via GP-based FDA;
\item Modeling temporal variability of stellar spectra due to stellar activity (starspots, plage regions, faculae) by combining FDA and bivariate function approximation using empirical separable expansions (i.e., the HSD described in \S~\ref{sec:FDA});
\item Constructing a low-dimensional empirical model for galaxy spectra by modeling SDSS galaxy spectra with a combination of FDA and manifold learning methods.
\end{itemize}
In the next section we describe the galaxy spectra application, an example of univariate FDA with an innovative model architecture that combines manifold learning with more conventional FDA approaches.
This work provides an example of fruitful interaction between machine learning and FDA.
In the subsequent section we describe the SN~Ia problem as an example of second-generation Bayesian FDA in astronomy.

\section{Modeling the population of galaxy spectra}
\label{sec:SEDPop}

Galaxies have diverse optical spectra%
\footnote{In this section, ``optical'' refers to light spanning visible wavelengths, and also the adjacent near-ultraviolet and near-infrared regions accessible to ground-based observatories.}
(also called spectral energy distributions, or SEDs), reflecting diversity in their stellar populations (which vary in age, mass distribution, and chemical composition), absorption and scattering by dust and gas, and the absence or presence of an active galactic nucleus (AGN), a supermassive central black hole contributing optical emission from an accretion disk and jet formed as surrounding gas is consumed.
Physical models for a single galaxy spectrum predict the emission from stellar populations by positing distributions for the masses and compositions of stars at birth, and a distribution of birth times, and then computing the SED as a function of age by accounting for stellar evolution.
Such models typically have a dozen or more parameters and they are computationally expensive (recent work is accelerating them via emulation with machine learning algorithms).
Fits to low- and modest-resolution spectral data typically exhibit significant parameter degeneracy (nonidentifiability), exacerbated by the possible presence of an AGN component (which can mimic the contribution of some stellar components) \cite{P+22-ArtSEDFit}.

It would be extremely useful to have an accurate and fast low-dimensional model for galaxy SEDs.
Such a model probably would not completely replace physical models for the purpose of estimating physical parameters for particular galaxies.
But a fast, low-dimensional model would be very useful for simulating galaxy populations (for the purpose of developing and testing large-scale data analysis pipelines for upcoming surveys).
And such a model could be crucial for photometric redshift estimation, where astronomers estimate how much a galaxy's spectrum has been redshifted by cosmological expansion, using only a few bands of photometric data as a crude spectrum measurement for simultaneous estimation of a galaxy's spectrum and the amount of redshift.
Redshift serves as a proxy for distance, a crucial galaxy parameter for cosmology.
Achieving the science goals of upcoming surveys requires accurate and precise photometric redshift estimation for hundreds of millions of galaxies.
Current physical models are too complex and slow for this purpose.
Black-box machine learning algorithms are not accurate enough.
Photometric redshift estimation using an accurate and fast low-dimensional empirical SED model is a promising approach.

\begin{figure}[ht]  
\begin{adjustwidth}{-.5\extralength}{0cm}
\includegraphics[width=15.5 cm]{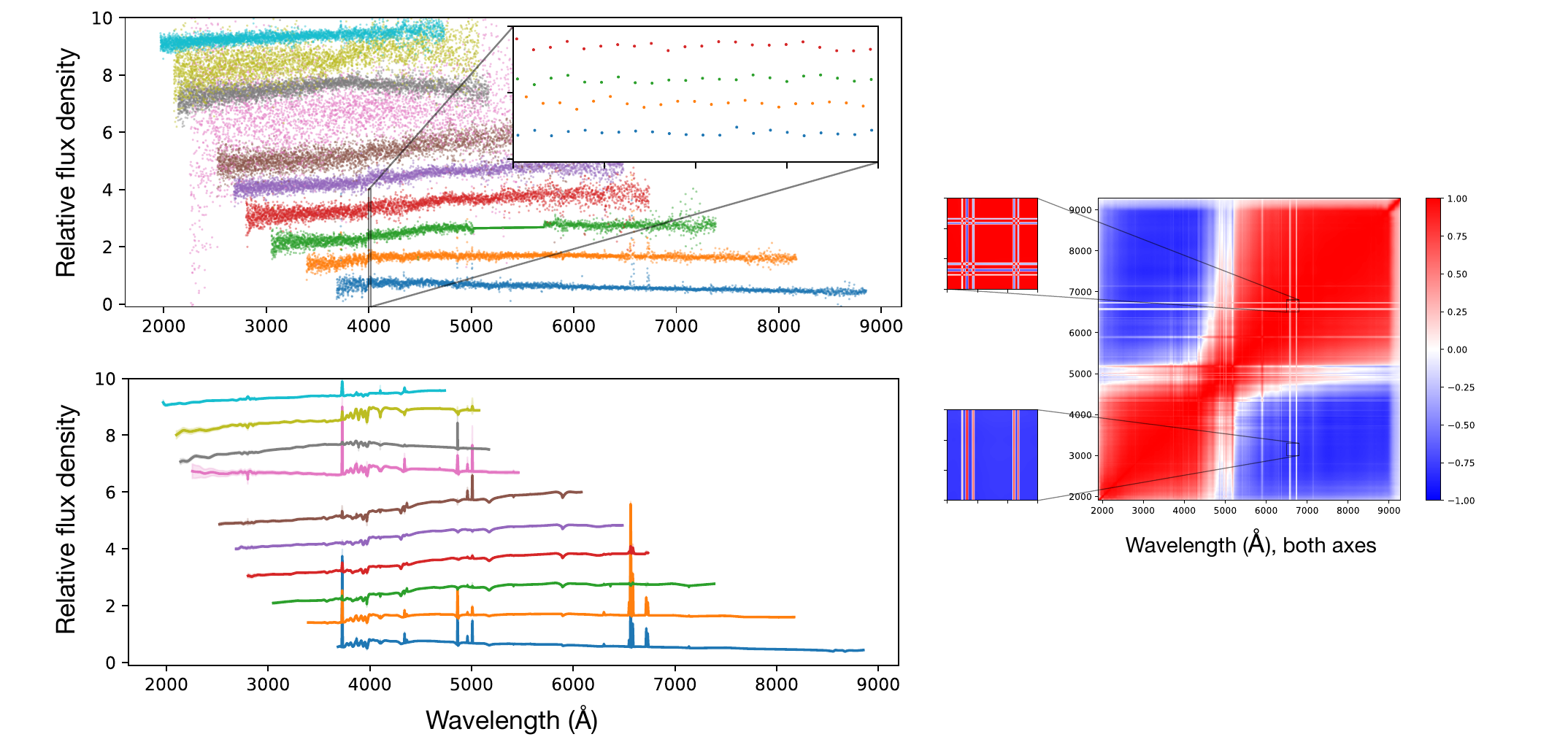}
\caption{
SnL model for the population of SDSS galaxy spectra.
\emph{Top:} Measurements of 10 SEDs shown (as dots colored for each SED).
\emph{Right:} Covariance function for GP SED population model.
\emph{Bottom:} Denoised and interpolated SEDs corresponding to top panel, with error bands (quite narrow for most SEDs; most perceptible for the pink SED). 
\label{fig:SnL}}
\end{adjustwidth}
\end{figure}

\begin{figure}[ht]  
\begin{adjustwidth}{-.5\extralength}{0cm}
\includegraphics[width=15.5 cm]{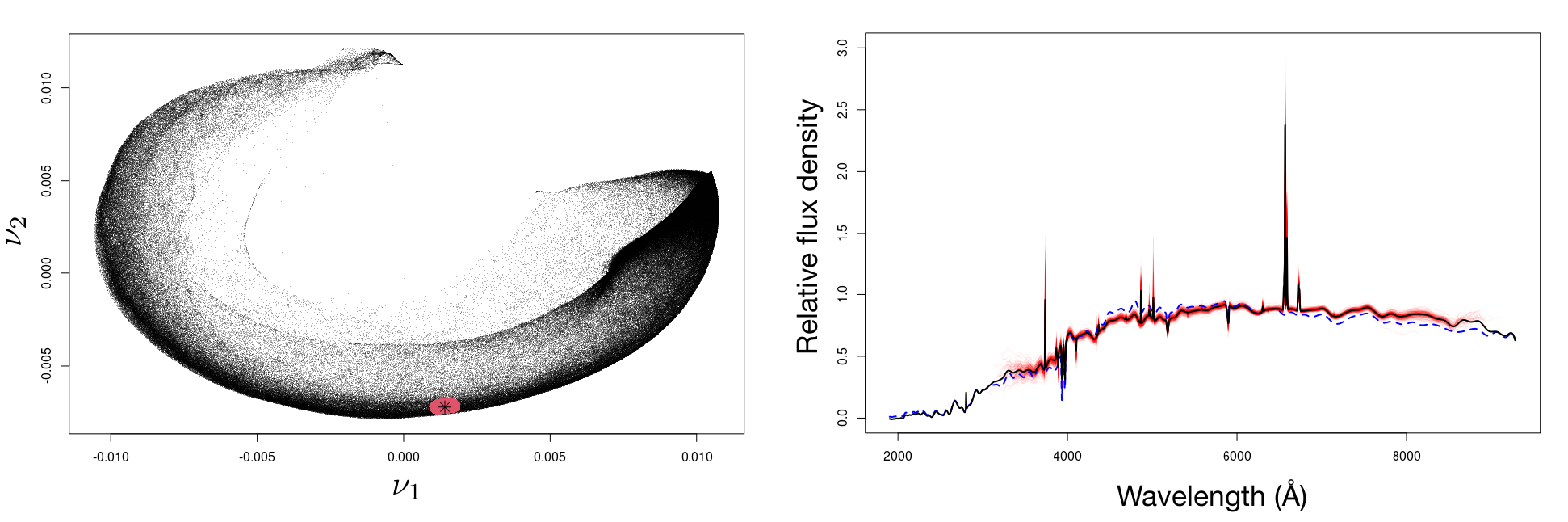}
\caption{
\emph{Left:} PHATE-based manifold for SDSS SEDs; 2D projection of 3D manifold shown, with manifold coordinates $(\nu_1,\nu_2)$.
\emph{Right:} Example predicted SEDs from inverse map.
Black curve shows SED corresponding to manifold coordinates shown by star in left panel.
Red curves span show SEDs spanning the red region containing the star in left panel.
\label{fig:ISM}}
\end{adjustwidth}
\end{figure}

We have constructed such a SED model using nearly 800,000 calibrated high-resolution SEDs observed by SDSS, with known redshifts (from measuring the locations of Doppler-shifted spectral line features).
Our framework has three components whose linked operation is depicted in Figures \ref{fig:SnL} and \ref{fig:ISM}:
\begin{itemize}
\item Denoising and interpolation of noisy measurements to the underlying continuous spectra (estimated using GP-based FDA).
\item Embedding of the continuous spectra to a low-dimensional space that preserves similarity of spectra (via manifold learning).
\item Construction of a map back from the low-dimensional space to full, continuous spectra (via function-on-vector regression).
This component is the final low-dimensional empirical SED model.
\end{itemize}

\unskip

The top panel in \cfig{fig:SnL} shows calibrated SED data for a small selection of SEDs, shifted to the galaxy rest frame using each galaxy's known redshift.
The SED data are not aligned, have noise at varying levels, and many spectra have gaps (e.g., the horizontal line in the green SED).
We model the collection of SEDs using a B-spline basis for the continuum, and Gaussian functions for the lines, with the basis coefficients and line effective areas drawn from a high-dimensional multivariate normal distribution \cite{K+23-SnL}.
This \emph{Splines-n-Lines} (SnL) model corresponds to a nonstationary degenerate Gaussian process (GP) model for the SED population, with a complex mean SED and a nonstationary, richly structured covariance matrix (specified by thousands of parameters).%
\footnote{A GP is degenerate if its covariance function has a finite number of non-zero eigenvalues.
This implies a kind of finite dimensionality of the process:
If one evaluates a sample path (function) from the process at more points than there are parameters specifying the process (here, values at spline nodes), then the covariance matrix for the values of the function will not have full rank.
Effectively, specifying values of the sampled function at some finite number of points (equal in number to the number of spline nodes) determines its behavior everywhere.}
The inferred covariance function is depicted in the right panel (as a matrix on a finite wavelength grid); it displays how parts of the spectra are related to each other, in particular, how line strengths depend on parts of the continuum.
For each SED, SnL estimates a denoised predicted spectrum---the curves in the lower panel---with uncertainties (e.g., a pointwise confidence band).
Notably, it successfully interpolates and moderately extrapolates incomplete SEDs, producing an ensemble of continuous SED estimates with large overlapping wavelength coverage.
SnL replaces ad hoc methods for putting SEDs on a common wavelength grid, e.g., via PCA on a grid of interpolated values.
It provides functional (gridless) representations, explicitly models the connection between lines and the continuum spectrum, and provides uncertainties across the estimated spectrum. 

\unskip

We next use the SnL-produced (gridless) SEDs as inputs for a manifold learning algorithm, seeking a low-dimensional representation.
Many manifold learning algorithms use a matrix of pairwise similarities to construct a low-dimensional relabeling of the data that approximately preserves distances.
We use the SnL-produced SEDS to compute a pairwise SED similarity matrix using cosine similarity.
The similarity matrix serves as input to the PHATE manifold learning algorithm, chosen for its ability to both preserve continuous, sequence-like structure in the data, and reveal clusters \cite{M+19-PHATE}.
PHATE identifies a 3D manifold in a learned SED manifold coordinate space denoted $\vec\nu = (\nu_1,\nu_2,\nu_3)$, shown in a 2D projection in the left panel of \cfig{fig:ISM}.
SEDs occupy a thin shell-like structure in the embedding space.

Finally, we build the final parametric empirical SED model, using the $\vec\nu$ coordinates as SED parameters, by inverting the map from SEDs to the manifold using a new function-on-vector regression algorithm.
It uses thin-plate splines as basis functions in $\vec\nu$ space to map back to the SnL spline and line coefficients.
The right panel shows the SED (black curve) corresponding to the indicated point in $\vec\nu$ space, along with several nearby SEDs (red) from the red region of the SED manifold.
Moving over the entire space smoothly generates SEDs spanning the full diversity seen by SDSS.

A population distribution in $\vec\nu$ space can be used to simulate SEDs without the discreteness imposed by current methods that simply draw from a finite set of prototypical SEDs.
A \photoz\ estimation algorithm can be built using the inverse map as a 3-parameter model to fit photometric data via ``synthetic photometry''---using a predicted SED to predict the photometry by computing functionals of the spectrum using known photometric filter response functions.

\section{Spectrotemporal Modeling of Type Ia supernova light curves and spectra}
\label{sec:SNIa}

Type Ia supernovae (SNe~Ia) are bright, cosmic explosions of white dwarf stars that are observed at great cosmological distances (and thus at much earlier times in cosmic history).
Estimates of their distances, along with the redshifts of the galaxies in which they occur, can thus be used to trace the history of the expansion of the universe---quantified by the current expansion rate (the Hubble constant, $H_0$), and its rate of change (``cosmic acceleration'')---and constrain the parameters of the cosmological components posited to influence it (e.g. dark matter and dark energy) \cite{Riess22-H0, Freedman25, Vincenzi24}.
To do this, astronomers use SNe~Ia as \emph{standardizable candles}.
With hypothetical \emph{standard candles}, the peak luminosities of the explosions would be the same across all objects, therefore the relative distances of the objects could be determined from measurements of their apparent brightnesses on Earth.
However, SNe~Ia do not all have exactly the same peak luminosity, but there are empirical correlations of their luminosities with observable properties of their multi-wavelength light curves, e.g. their temporal evolution, as measured by their ``width'' or decline rate after peak, and their colors (relative brightness across wavelength) \cite{Phillips93, Tripp98, Mandel17}.
These empirically-determined correlations enable astronomers to model the luminosities of individual SNe~Ia (i.e., standardize them), so that their distances can then be determined from their apparent brightnesses.

Several characteristics of the latent physical processes and the observational processes that produce the data present particular challenges for statistical modeling and inference. The data are multiband light curves---multivariate time series of apparent brightness measurements in different filters covering neighboring wavelength ranges.  These time series are often sparsely and irregularly sampled, and asynchronously sampled across different filters, depending on a particular survey's observing strategy, complicating joint modeling of the multiband data. The peak brightness of a SN light curve is a conventional parameter of interest, but due to sparsity of the data there is seldom direct measurement of the brightness of a SN Ia in any filter exactly at the light curve peak, and in many cases supernovae may be detected only after peak brightness. Because of the redshifting effect of the expansion of the universe on supernova spectral energy distributions, the filters in our telescopes (in the \emph{observer frame}, on the ground or in space near the Earth) collect light corresponding to different wavelength ranges in the \emph{rest frames} of SNe~Ia that are at different redshifts (a photon observed at wavelength $\lambda_\text{obs}$ was emitted at the shorter (more blue) rest frame wavelength $\lambda = \lambda_\text{obs} / (1+z)$). Additionally, the observed temporal evolution is slower than in the SN~Ia rest-frame due to cosmological time-dilation (a rest-frame time interval $\delta t$ is observed to span $\delta t_\text{obs} = \delta t \times (1+z)$). Finally, the observed dependence of an dynamic SED on wavelength and time arises from two types of astrophysical processes: processes \emph{intrinsic} to the supernovae related to the progenitor star and explosion physics (light generation processes), and \emph{extrinsic} processes that cause dimming and reddening of the light due to interstellar dust and gas along the line-of-sight to the observer (light propagation processes).  Ignoring the dimming due to propagation effects will lead to mistakenly large inferred SN distances.

To optimally use SNe~Ia as cosmological distances indicators, we model the rest-frame SEDs as a population of continuous functions of both time and wavelength, in a framework that can simultaneously account for all of these complexities. Explicitly modeling the rest-frame SN~Ia dynamic SEDs enables us to predict the multi-wavelength time series in the observer frame via ``forward modeling'' of physical and observational effects. By accounting for the redshift effects on each SN~Ia, this approach can estimate and compare the temporal and spectral characteristics across an ensemble of supernovae spanning a large range of redshifts, by mapping the observations to common rest-frame wavelength and time ranges. Since the variations and correlations of SED functions over time and wavelength must be learned over a \emph{population} of SNe~Ia, this problem is ideally addressed from a Bayesian FDA perspective. This is the approach taken by BayeSN.

BayeSN is a hierarchical Bayesian spectrotemporal model for observations of the time-dependent SED functions of SNe~Ia spanning the optical to near-infrared (NIR) wavelength regime \cite{Thorp21, M+22-HBSED-SNIa, Grayling24}\footnote{It builds on earlier hierarchical Bayesian models for the multivariate time-series of the integrated light of SNe~Ia within a discrete set of filters, i.e. continuous in time but discrete in wavelength \cite{Mandel09, Mandel11,Mandel14}.}.  The hierarchical model describes the distribution of observed dynamic SED functions as arising from the interaction of two physical populations: the population of intrinsic SN Ia dynamic SEDs arising from diversity of the properties of the progenitor stars and explosion physics, and the population of SN host galaxy dust properties that determine propagation effects. Because of dust's strong but smooth dimming effect as a function of wavelength, modeling the wide optical-to-NIR wavelength range enables the model to better disentangle the intrinsic effects from the dust effects.

When used to analyze the multi-wavelength time series data of a large sample of SNe~Ia, the hierarchical model fits the light curves of individual SNe~Ia, while learning the statistical structure of a population of SN Ia SED functions $S(t,\lambda)$.  The learned statistical structure (represented by the population hyperparameters of the hierarchical model) encodes the correlations across time and wavelength needed to optimally determine the distance to any individual SNe~Ia from its photometric time series.

\begin{figure}[ht]  
\begin{adjustwidth}{-.5\extralength}{0cm}
\includegraphics[width=15.5 cm]{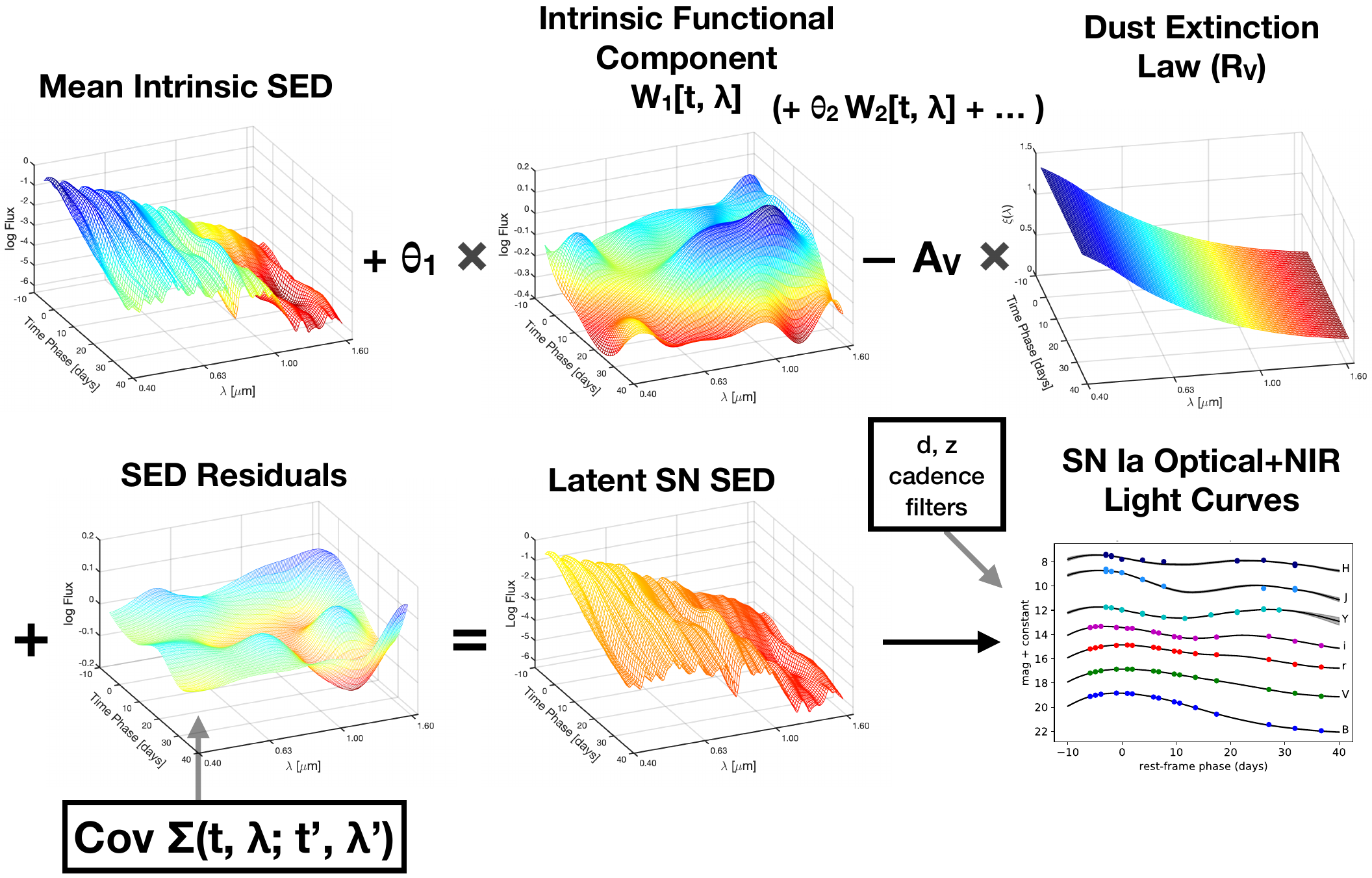}
\caption{Schematic of the \textsc{BayeSN} forward generative model for light curve (time series) data of a single SN. The log SED, a continuous function of rest-frame time and wavelength, comprises a mean intrinsic SED function modified by intrinsic functional principal components scaled by latent SED shape parameters, and then dimmed by the host galaxy dust law (a function of wavelength), parametrized by a shape parameter $R_V$, and scaled by the amount of dust $A_V$. Intrinsic variations not captured by these major modes are modeled by residual SED functions whose statistical properties across time and wavelength are captured by a covariance function.  The resulting latent rest-frame SED undergoes the effects of distance, redshifting, and time dilation, integration of the brightness under the observer's filter transmission functions, the survey cadence, and measurement error to yield the observed multi-band  time series (light curves) of a SN Ia. Adapted from \citet{M+22-HBSED-SNIa}.
\label{fig:bayesn_fig1}}
\end{adjustwidth}
\end{figure}

The key equation of BayeSN expresses the logarithm of the SED of an SN~Ia indexed by $s$, a continuous function of the rest-frame time $t$ and wavelength $\lambda$, as a linear combination of terms in the manner of a mixed-effects model:
\begin{equation}\label{eqn:M22_logsed}
\begin{split}
-2.5 \log_{10} \left(\frac{S_s(t, \lambda)}{S_0(t,\lambda)}\right) &= M_0 + W_0(t,\lambda) \\
&\quad {} + \delta M_s + \left[ \sum_{k=1}^K \theta_k^s \, W_k(t,\lambda) \right] + \epsilon_s(t,\lambda) \\
&\quad {} + A_{V}^s \, \xi(\lambda; R_V^{(s)}).
\end{split}
\end{equation}
The left-hand-side is the logarithm of the ratio of the SN-specific SED for SN $s$ to a global, average spectrotemporal template $S_0(t,\lambda)$ \cite{hsiao07, hsiaothesis}. 
The right hand side is divided into three parts: terms shared across all SNe (fixed effects), and two supernova-specific parts ($s$-dependent random effects), one describing intrinsic SED diversity, and one describing diversity in host galaxy dust properties.
The terms on on the RHS first line describe a global normalization term and a global time- and wavelength-dependent average intrinsic dynamic SED template.
The terms on the second line describe SED effects particular to each SN $s$.
The $\delta M_s$ term describes a time- and wavelength-independent intrinsic ``scatter'' of SN~Ia luminosities.
The second term decomposes the spectrotemporal behavior into the sum of functional principal components (FPCs) $W_k(t,\lambda)$ multiplied by SN-specific coefficients $\theta_k^s$. 
The remainder intrinsic, residual spectrotemporal variation not captured by the FPCs is modelling by residual functions $\epsilon_s(t,\lambda)$, drawn from a zero-mean stochastic process with a flexible covariance function over time and wavelength (see below for the construction).  
Finally, the term on the third line describes the effect of the host galaxy's dust on the SED: $A_V^s$ is proportional to the amount of dust along the line-of-sight to SN $s$, $\xi(\lambda, R_V^{(s)})$ describes the dimming effect as a function of wavelength  (called the dust law), and $R_V^{(s)}$ is a parameter controlling the shape of this function, and can also be different for each SN $s$.
This forward model is depicted graphically in Figure \ref{fig:bayesn_fig1}.
The SED resulting from these effects is then dimmed according to distance (using the cosmological generalization of the inverse-square law), redshifted and time-dilated using the SN redshift, and then used to predict observed fluxes, using a telescope's filter response functions and a survey's observing cadence (sample times).

\unskip

While Eq. \ref{eqn:M22_logsed} describes the SN Ia SED for an individual supernova, the hierarchical model is completed by specifying population distributions over each SN-specific parameter or function, and hyperpriors over population hyperparameters and global functions (specified in detail in \citet{M+22-HBSED-SNIa}).  In particular, prior distributions are assigned to the coefficients $\theta_k^s$ and the FPCs describing the intrinsic effects, so that they are effectively learned using a two-dimensional functional generalization of Bayesian PCA \cite{tipping99, bishop99}, simultaneously with inferring the individual dust effects and their population distribution. The FPCs $W_k(t,\lambda_r)$ and the residual functions $\epsilon_s(t,\lambda_r)$ are modeled as splines over rest-frame time and wavelength, described by their values on a grid of spline nodes, which are learned by training the hierarchical model on a SN sample. The joint population distribution of the residual functions' nodes is modeled as a multivariate Gaussian distribution, whose covariance matrix is also learned. Since the splines are linear in the node values, this implies that the residual functions themselves are realizations of a (degenerate) Gaussian process with a non-stationary covariance function over time and wavelength implied by the learned covariance matrix.

\begin{figure}[ht]  
\begin{adjustwidth}{-.5\extralength}{0cm}
\includegraphics[width=15.5 cm]{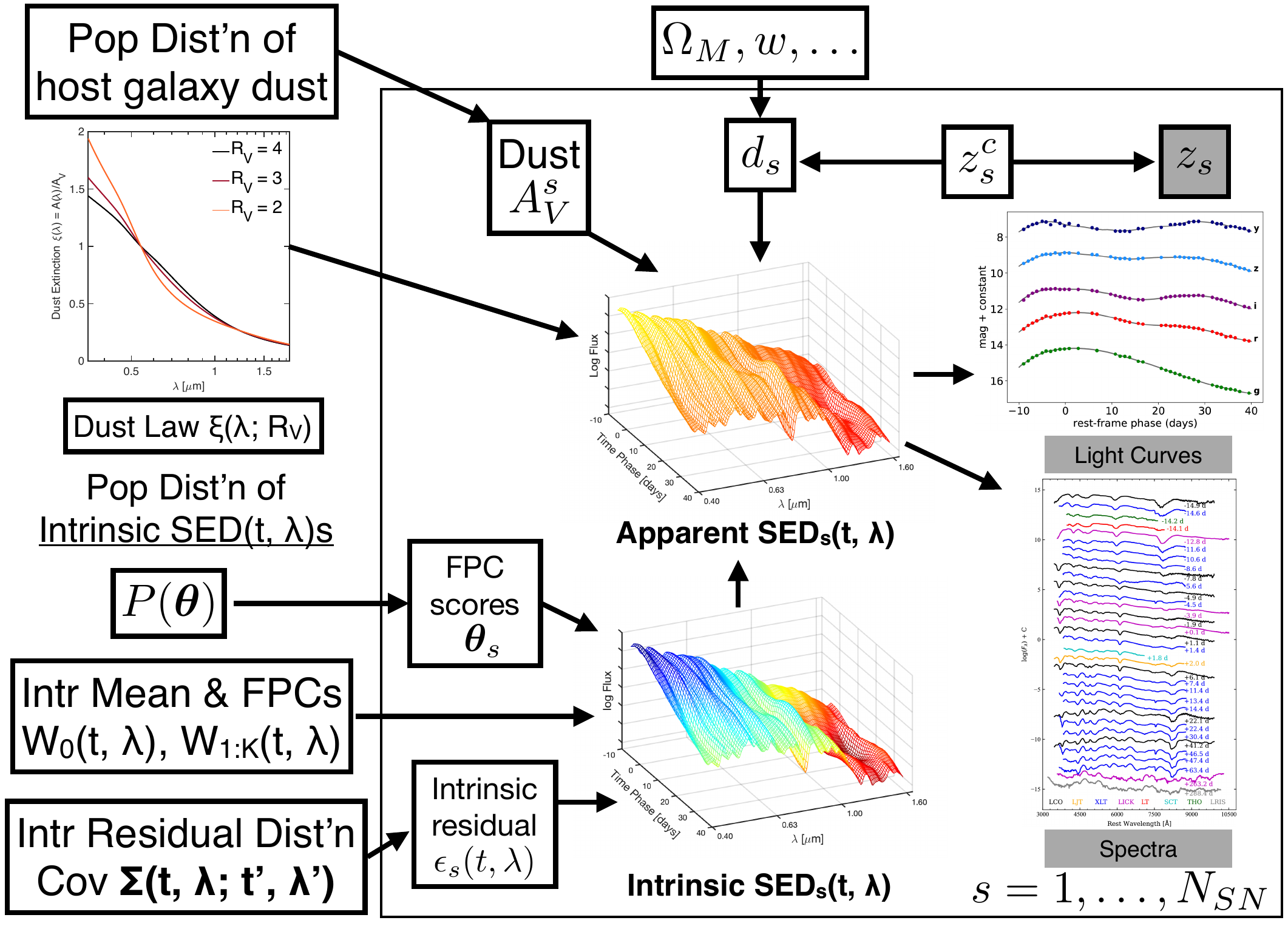}
\caption{Probabilistic graphical model depicting the hierarchical \textsc{BayeSN} SED model for SN Ia photometric time series (light curve) and spectroscopic time series data. Each open box represents a set of unknown parameters or hyperparameters, each grey-shaded box represents observed data, and the arrows indicate relations of conditional probability.  Boxes within the plate, labelled $s = 1,\ldots,N_\text{SN}$, represent latent variables or functions specific to each SN $s$ within a sample of $N_\text{SN}$ SNe~Ia, whereas boxes outside the plate represent global or population functions or hyperparameters. The hierarchical posterior distribution estimates the unknown individual latent variables and global hyperparameters conditional on the observed data of the entire SN Ia sample. Insets show photometric time series (light curves) \cite{Ward23} and spectroscopic time series \cite{Iskander25} data of SN 2021hpr. Adapted from \citet{M+22-HBSED-SNIa}.
\label{fig:bayesn_fig2}}
\end{adjustwidth}
\end{figure}   
\unskip

Figure \ref{fig:bayesn_fig2} shows a directed acyclic graph describing the hierarchical model's joint distribution for the data, latent variables of individual SNe~Ia, and hyperparameters of the intrinsic and dust population components.
Current versions of BayeSN were trained primarily on the photometric light curves (upper right) of SN~Ia training sets.
However, the ultimate goal is a \emph{multimodal} hierarchical Bayesian model that learns simultaneously from both photometric and spectroscopic time series (lower right) of SNe~Ia, as the latter will improve the spectral features of the learned FPCs on fine wavelength scales (Grayling et al., in prep.).
These two modalities of data both probe the latent SED, but give complementary information.
Photometry provides time series of low-resolution information (in wavelength) about the underlying SED. 
We model these as noisy measurements of integrals of the SED under the filter's transmission function.
Spectra are high-resolution slices in wavelength through an SN's SED at particular times, but are costly to obtain and so are usually sampled more sparsely in time.
(In the era of LSST, the vast majority of SNe~Ia discovered will not have any spectra obtained).
Principled hierarchical modeling and inference of populations of SED functions with the multi-resolution information contained in photometric and spectroscopic data presents unique challenges for Bayesian FDA. \citet{boyd25} presented a recent example of simultaneous hierarchical Bayesian modeling of photometry and spectra in the case of static sources (white dwarf stars) to precisely calibrate an all-sky network of state-of-the-art spectrophotometric standards.

Training the BayeSN hierarchical model entails estimating all population hyperparameters and latent variables of individual SNe of a training sample, which currently can total of order $\sim 10^4$ scalar parameters, and presents  computational scalability challenges. \citet{Grayling24,Grayling26b} presented an efficient Hamiltonian Monte Carlo (HMC) code that utilizes automatic differentiation and GPU-acceleration to train BayeSN (by sampling the full hierarchical posterior distribution) on light curve samples of $\sim 10^3$ SNe~Ia in less than an hour\footnote{\url{https://github.com/bayesn/bayesn}}. With the hierarchical model trained (hyperparameters estimated), the model can be deployed to fit light curves of new individual SNe Ia, also using HMC. Further computational speed-ups have been achieved by applying variational methods to approximate the posterior for the latter inference problem \cite{Uzsoy24}, enabling scaling to much larger datasets. However, applying variational methods to full hierarchical inference remains a challenge. Further scalability gains may be realized via simulation-based inference methods that avoid direct evaluation of an explicit likelihood, and learn a mapping between  hyperparameters of interest and the samples of SN~Ia time series using forward simulations and neural networks. For example, BayeSN has been incorporated into a forward simulator to train a neural ratio estimator to perform marginal posterior inference of population hyperparameters directly from the collection of time series of a sample of SNe~Ia \cite{Karchev24} . Such simulation-based inference approaches show promise for incorporating realistic selection effects into the population and cosmological inference \cite{Boyd24, Karchev25, Boyd26}.

The BayeSN hierarchical SED model has been applied to many astrophysical and cosmological studies of SNe~Ia. Because it models the SN Ia SEDs as arising from the combination of two population distributions (intrinsic and dust), it has been deployed to investigate and constrain the possible dependencies of the component population distributions on the properties of the SN~Ia host galaxies (e.g. their stellar masses) and/or their possible evolution with redshift \cite{Thorp21, Thorp24, Grayling24, Grayling25, Ginolin26}, which are major systematic uncertainties in contemporary supernova cosmology analyses \cite[e.g.][]{Vincenzi24}. \citet{Thorp22} demonstrated that the fitting of light curves of individual SNe~Ia separately to determine the population distribution of dust law shapes, and then aggregating the individual estimates, leads to an overestimate of the population dispersion; instead a hierarchical SED analysis produced shrinkage estimates that were more accurate.  BayeSN has been applied to a novel hierarchical SED analysis of supernova siblings---multiple SNe~Ia located in the same galaxy and therefore at the same distance \cite{Ward23}. BayeSN has also been used to jointly analyze optical and NIR light curves of relatively nearby low-redshift SNe~Ia calibrated by Cepheids (\S~\ref{sec:Tab2Func}) for the determination of the Hubble constant \cite{Dhawan23}, to analyze novel Hubble Space Telescope NIR observations of cosmologically-distant, high-redshift SNe Ia used to constrain the properties of dark energy \cite{Jones22, Thorp24}, and to analyze James Webb Space Telescope NIR observations of very high-redshift SNe~Ia to estimate the Hubble constant using strong gravitational lensing time delays \cite{Pierel24, Pascale25,Grayling26a} and to test their intrinsic luminosity evolution over cosmic time \cite{Pierel25-LumEvo}.

\section{Closing remarks}
\label{sec:Closing}

Current and emerging astronomical instruments place astronomy in the midst of a veritable flood of functional data, across the electromagnetic spectrum.
Dedicated exoplanet surveys are producing dynamic spectra at extremely high spectral resolution for thousands of stars; successful detection of Earth-like planets will require careful modeling of stellar activity signals in the data that can mask or mimic planet signals.
Orbiting X-ray and gamma-ray instruments are producing time series, spectra, and dynamic spectra for thousands of sources.
The Dark Energy Spectroscopic Instrument (DESI) is producing a growing sample of spectrophotometrically calibrated galaxy spectra that dwarfs the SDSS sample.
The Rubin Observatory's LSST survey, mentioned in the Introduction, has begun to turn the flood into a torrent; started in June 2026, its ten-year survey will produce multiband time series and images of many billions of stars, galaxies, and transient phenomena.

These developments surely motivate broader adoption of FDA in astronomy, with Bayesian FDA particularly well suited to handling the complexities of irregular and asynchronous sampling, nontrivial measurement error, and selection effects.
They also point to opportunities for FDA statisticians, not only to bring FDA insights into astronomy, but also to draw on software tools that astronomers are developing for access and manipulation of  huge functional datasets.
An example is the \npandas\ project \cite{M+25-NP+LSDB}, developing a generalization of the \rlang\ dataframe optimized for handling irregularly and asynchronously sample multivariate functional data in \python.


The immense scale of emerging datasets in astronomy has motivated increasing interest in deep learning methods, which typically scale better than nonparametric and semiparametric statistical methods \cite{T26-DLInAstro}.
In many applications, these methods are essentially modeling and predicting measurements of populations of functions (e.g., light curves, spectra, or images). This is particularly the case with multimodal foundation models, deep learning architectures trained on large ensembles of diverse data with the goal of learning representations useful for diverse downstream tasks (typically after task-specific post-training).
The learned representations are often implicitly descriptions of function populations.
Deep learning methods used in astronomy typically have input layers that presume data are evenly sampled on a shared sample or pixel grid, essentially treating the data as fixed-dimension vector data (though some layers within a deep network may implicitly acknowledge the continuous function nature of the underlying objects to be modeled).
These methods often struggle with sound uncertainty quantification.
These issues motivate research seeking to connect FDA and machine learning methods, uniting the statistical foundations of FDA with tools from machine learning for flexible discovery and representation of high-dimensional structure in large datasets.
It is notable that application of Bayesian FDA to the two areas highlighted above---modeling galaxy SEDs, and modeling SN~Ia light curves---brings together statistical tools from the FDA literature (splines, hierarchical modeling) with tools from machine learning (Gaussian processes, kernel methods, manifold learning).
Specifically connecting FDA to \emph{deep} learning is a largely open research area (see \cite{W+24-FDA+DNNs} for a recent review of the limited existing literature).
The combination of large functional datasets, compelling scientific questions, and the public accessiblity of the data (without proprietariness, privacy, and security issues that arise in other application areas) makes astronomy an ideal setting for 
exploring the potential merger of the (Bayesian) FDA perspective and the powerful tools of deep learning.


\vspace{6pt} 


\authorcontributions{Conceptualization, T.L., T.B., M.G., D.R., and K.M.; methodology, D.K., D.R., T.L., T.B., K.M., and K.T.; software, D.K., T.L., K.T., and K.M.; writing---original draft preparation, T.L. and K.M.; writing---review and editing, T.B., M.G., D.K., D.R., K.T.; supervision, T.L., D.R., T.B., and K.M. 
All authors have read and agreed to the published version of the manuscript.
}

\funding{The authors' research described here was supported by NSF grants AST-1814840, AST-2206339, and DMS-2210790 at Cornell University, by NSF grants AST-1814778 and AST-2206341 at Johns Hopkins University, and by NSF grant AST-2206340 at Caltech.
Research at the University of Cambridge was supported by the European Union’s Horizon 2020 research and innovation programme under European Research Council Grant Agreement No.~101002652 (PI K. Mandel) and Marie Skłodowska-Curie Grant Agreement No.~873089.
}


\acknowledgments{We thank Martin Hendry and Daniel Kowal for contributing to the development of a proof-of-concept Cepheid light curve FDA model briefly mentioned in \S~\ref{sec:AstroFDA};
this was the first foray into FDA for astronomy for T.L.\ and D.R..
We thank Eduardo Cristo for assistance with the SOAPv4 stellar activity simulation software.
We thank Anela Wang for assistance with proof-of-concept work on light curve classification with (multitask) Gaussian process FDA, briefly mentioned in S~\ref{sec:AstroFDA}.
}


\conflictsofinterest{The authors declare no conflicts of interest.}

\begin{adjustwidth}{-\extralength}{0cm}

\reftitle{References}


\bibliography{AstroFDARefs}


\PublishersNote{}
\end{adjustwidth}
\end{document}